\documentstyle[psfig,lscape]{mn}
\def\la{Ly$\alpha$}
\def\etal{et~al.}
\def\spose#1{\hbox to 0pt{#1\hss}}
\def\lta{\mathrel{\spose{\lower 3pt\hbox{$\mathchar"218$}}
     \raise 2.0pt\hbox{$\mathchar"13C$}}}
\def\gta{\mathrel{\spose{\lower 3pt\hbox{$\mathchar"218$}}
     \raise 2.0pt\hbox{$\mathchar"13E$}}}
\def\Ha{H$\alpha$}
\def\Hb{H$\beta$}
\def\Hc{H$\gamma$}
\def\Hd{H$\delta$}

\def\clean{{\sc clean}}

\def\aips{{\sc aips}}

\def\iraf{{\sc iraf}}
\def\noao{{\sc noao}}

\def\kms{\,km\,s\,$^{-1}$}
\def\Ho50{$H_0 = 50$km\,s$^{-1}$\,Mpc$^{-1}$}

\title[The most powerful equatorial radio sources]{A 98\%
spectroscopically complete sample of the most powerful equatorial radio
sources at 408\,MHz}

\author[P.~N.~Best \etal]{P.~N.~Best,$^1$\thanks{Email:
pbest@strw.leidenuniv.nl} H.~J.~A.~R\"ottgering$^1$ and
M. D. Lehnert$^2$\\ 
$^1$ Sterrewacht Leiden, Postbus 9513, 2300 RA Leiden, the Netherlands\\ 
$^2$ Max-Planck-Institut f{\"u}r extraterrestrische Physik, P.O.Box 1603,
87450 Garching, Germany}

\pagerange{\pageref{firstpage}--\pageref{lastpage}}
\pubyear{1999}
\begin{document}
\label{firstpage}
\setcounter{topnumber}{3}
\setcounter{dbltopnumber}{4}

\maketitle

\begin{abstract}
\noindent A new sample of very powerful radio galaxies is defined from the
Molonglo Reference Catalogue, according to the criteria $S_{\rm 408 MHz} >
5$\,Jy, $-30^{\circ} \le \delta \le 10^{\circ}$ and $|b| \ge
10^{\circ}$. The sample is selected to have similar properties to the
northern 3CR revised sample, and to be visible to a combination of
existing northern telescopes such as the Very Large Array radio
interferometer and large southern hemisphere telescope facilities. The
sample contains 178 sources, of which spectroscopic redshifts are
available in the literature for 128. For the remaining 50 sources, new
radio imaging, optical imaging and spectroscopic observations are
presented to identify the host galaxies and determine their
redshifts. With these new observations the total sample is 100\% optically
identified and redshifts are available for 174 (98\%) of the sources. The
sample consists of one starburst galaxy, one Seyfert galaxy, 127 radio
galaxies and 49 quasars. Basic properties of the sample, such as the
distributions of the quasar and radio galaxy populations in redshift and
their locations on the radio power versus linear size ($P-D$) diagram,
show no significant differences from the revised 3CR sample. The
equatorial location and the high spectroscopic completeness of this sample
make it a valuable resource for detailed studies of the nature and
environments of these important objects with the new generation of
southern hemisphere telescopes.
\end{abstract}


\section{Introduction}

Radio sources are unique cosmological probes, with importance for
understanding the physics of active galactic nuclei, for studying the
relationship between the radio source and its surrounding environment, for
probing high redshift proto--cluster environments, and for defining
complete samples of galaxies for studies of stellar populations at early
epochs. The revised 3CR sample (Laing, Riley \& Longair 1983; hereafter
LRL)\nocite{lai83} contains the brightest extragalactic radio sources in
the northern sky selected at 178\,MHz; the host galaxies of these radio
sources are predominantly giant elliptical galaxies and lie at redshifts
out to $z \sim 2$. Scientifically, the revised 3CR sample has proven
exceedingly powerful since it is 100\% spectroscopically complete,
avoiding many selection biases inherent in less complete samples, and has
become established as the standard sample of bright low frequency selected
radio sources. The 3CR galaxies and quasars have been the subject of
extensive studies over a wide range of wavelengths leading to many
important discoveries, not least of which were the very tight relationship
between the infrared K--magnitudes and the redshifts of the radio galaxies
(e.g. Lilly and Longair 1984)\nocite{lil84a}, the discovery that the
optical and ultraviolet emission of the high redshift ($z \gta 0.6$) radio
galaxies is elongated and aligned along the direction of the radio axis
\cite{mcc87}, and the orientation--based unification schemes of radio
galaxies and radio loud quasars (e.g. Barthel 1989)\nocite{bar89}.

The new generation of large optical telescopes provides an
exciting new opportunity for very detailed studies of these important
objects and their environments, as has been proven by the results being
produced by the Keck telescope (e.g. Cimatti \etal\ 1996, Dey \etal\ 1997,
Dickinson 1997).\nocite{cim96,dey97,dic97a} Radio astronomy has, however,
historically been concentrated in the northern hemisphere, and there is
currently no large, spectroscopically complete sample of low frequency
selected radio sources equivalent to the 3CR sample for studies with
southern hemisphere telescopes such as the VLT and Gemini South. The
current paper aims to rectify this deficiency.

The layout of the paper is as follows. In Section~\ref{defin} the
selection criteria of the new sample are described. Details of the
observations that were carried out to provide optical identifications and
spectroscopic redshifts for those sources for which such data could not be
found in the literature are provided in Section~\ref{observe}. In
Section~\ref{eso}, the results of these observations are provided in the
form of radio maps, optical images and spectra of these sources. Tabulated
details of the resulting complete sample are compiled in
Section~\ref{details} and global properties of the sample are
investigated. Conclusions are summarised in Section~\ref{conc}. Values for
the cosmological parameters of $\Omega = 1$ and $H_0 = 50$\kms\,Mpc$^{-1}$
are assumed throughout the paper.

\section{Sample Definition}
\label{defin}

The basis dataset for our sample was the Molonglo Reference Catalogue
(MRC; Large \etal\ 1981)\nocite{lar81}, a catalogue of radio sources
selected at 408\,MHz in the region of sky $-85^{\circ} < \delta <
18.5^{\circ}$, $|b| \ge 3^{\circ}$, and essentially complete down to a
flux density limit of 1.0\,Jy at that frequency.  The low frequency
selection criterion of this catalogue, like that of the 3CR sample
(178\,MHz), selects radio sources primarily on the relatively steep
spectrum synchrotron emission of their extended radio lobes, rather than
on flat spectrum cores, jets and hotspots, and is therefore less subject
to Doppler boosting effects than samples selected at higher radio
frequencies.

The sample (hereafter the BRL sample) was drawn from the MRC according to
four criteria:

\begin{itemize}
\item They must have a flux density $S_{\rm 408 MHz} \ge 5$\,Jy.

\item They must lie in the region of sky $-30^{\circ} \le \delta \le
+10^{\circ}$.

\item They must lie away from the galactic plane, $|b| \ge 10^{\circ}$.

\item They must be associated with extragalactic hosts.
\end{itemize}

The first selection criterion is similar to the flux density limit of the
revised 3CR sample (LRL), $S_{\rm 178 MHz} > 10.9$\,Jy, for a typical
radio source with a radio spectral index $\alpha \sim 0.8$ ($S_{\nu}
\propto \nu^{-\alpha}$). The second criterion was made so that the sample
would be visible from both northern radio telescopes such as the VLA and
southern hemisphere telescopes such as the new large optical telescopes
(VLT, Gemini South) and the proposed Atacama Large Millimetre Array
(ALMA). The third criterion rejects most galactic objects and avoids the
regions of highest galactic extinction.

The first three selection criteria produced a sample of 183 entries in the
Molonglo Reference Catalogue. Of these, 0532$-$054 (M42, Orion) and
0539$-$019 were excluded on the basis of being galactic HII
regions. 0634$-$204 and 0634$-$206 appear as two separate entries in the
MRC catalogue, whereas they are in fact two parts of the same giant radio
source \cite{bau88}; these two entries were therefore merged as
0634$-$205. Similarly, the entries 1216+061A and 1216+061B are from the
same source, hereafter referred to as 1216+061 (e.g. see Formalont
1971)\nocite{for71}. Finally, the single catalogue entry 2126+073 (3C435)
is actually composed of two individual radio sources \cite{mcc89} neither
of which on its own is luminous enough to make it into the sample, and so
these were excluded. 0255+058 (3C75) is also composed of two distinct
sources, although overlapping and inseparable in terms of flux density
\cite{owe85}, but this entry is maintained within the sample since at
least one of the two sources must be sufficiently luminous that it would
have entered the sample on its own. These considerations led to a final
BRL sample of 178 sources.

Selected in this way the sample is complementary to many other radio
source samples which have been constructed or are under construction. The
northern declination limit of +10$^\circ$ corresponds to the southern
declination limit of the LRL sample. Bright sources more southerly than
the $-30^{\circ}$ declination limit will be included in a new southern
sample being prepared by Hunstead and collaborators. Between these three
samples, therefore, almost the entire sky (away from the galactic plane)
will be covered. The BRL sample further complements the MRC strip 1 Jansky
sample defined by McCarthy and collaborators (e.g. McCarthy \etal\
1996)\nocite{mcc96b}; the MRC strip, also composed of southern radio
sources from the MRC catalogue, is selected to be about a factor of five
lower in radio power than our sample and is currently $\gta 75\%$
spectroscopically complete. Due to its larger sky coverage, the BRL sample
provides almost a factor of 5 more radio sources at the highest flux
densities than the MRC strip, which is essential to provide large enough
numbers of the rare high power objects for studies of, for example, the
alignment effect or their clustering environments at high
redshifts. Combining the BRL sample and the MRC strip will allow
variations with radio power to be investigated. Finally, the new sample
provides a complement to samples selected at high radio frequencies, such
as that of Wall and Peacock \shortcite{wal85}, which contain far larger
fractions of flat spectrum sources and quasars than low frequency selected
samples, due to Doppler boosting effects. The Wall and Peacock sample
covers the whole sky away from the galactic plane ($|b| \ge 10^{\circ}$),
contains 233 sources brighter than 2\,Jy at 2.7\,GHz, and in the region of
sky $\delta < +10^{\circ}$ is over 90\% spectroscopically complete
(e.g. di Serego Alighieri \etal\ 1994)\nocite{dis94c}.

\section{Observations and Data Reduction}
\label{observe}

A literature search showed that spectroscopic redshifts were already
available for 128 of the 178 sources in the sample. Our observations
concentrated upon the remaining 50 sources, with the goal of producing a
spectroscopically complete sample.

\subsection{Radio Imaging}

The angular resolution of the observations comprising the Molonglo
Reference Catalogue is only about 160 arcseconds in right ascension and in
declination varies from about 170 arcseconds at $\delta = -30^{\circ}$ to
240 arcseconds at $\delta = +10^{\circ}$. In general these positional
uncertainties are too great to allow unambiguous identification of the
host radio galaxy or quasar; higher resolution radio data are essential.

Radio data with angular resolution of about 10 arcseconds or less were
extracted from the VLA archive for 27 of the 50 radio sources without
spectroscopic redshifts. Details of the array configurations and
frequencies of these archive data are provided in Table~\ref{obstab}. For
the remaining 23 sources new observations were made using the VLA during
filler time in September 1997 and June 1998 (see Table~\ref{obstab}).
These observations were single snapshot exposures of typically 5--minute
duration. Either 3C286 or 3C48 was observed during each run for primary
flux calibration, and accurate phase calibration was achieved by
observations of secondary calibrators within 10--15 degrees of the radio
galaxies.

Both the new data and that extracted from the archive were calibrated,
\clean ed, and further self--calibrated within the \aips\ package provided
by the National Radio Astronomy Observatory following standard reduction
procedures. The resultant radio maps are shown in Figures~1 to 50.

\subsection{3.6m EFOSC2 observations}

The optical imaging and spectroscopic observations were carried out
predominantly during two observing runs at the ESO 3.6m telescope, La
Silla, on 21-22 April 1998 and 20-21 November 1998. The EFOSC2 instrument
was used together with the 2048 by 2048 Loral CCD \#40 which, after
binning 2 by 2 pixels on read-out, provided a spatial scale of 0.315
arcsec per binned pixel.  The spectroscopic observations were taken
through a 2 arcsecond slit using the grism \#6 which provided a wavelength
coverage from 3860 to 8070\AA, a scale of 4.1\AA\ per binned pixel and a
spectral resolution of about 22\AA.

The observing technique used was to make first a short image of the field
of the radio source through the $R$--Bessel filter and use the VLA radio
map to identify the host galaxy or quasar. The telescope was then moved to
centre this object in the slit and a spectrum taken.  The spectral slit
position angle was left in the default east--west direction unless there
was good reason to do otherwise, for example if there were two candidate
host objects in which case they were both placed along the slit. The
duration of the spectral exposure was of between 5 and 20 minutes,
depending roughly upon the magnitude of the host object in the $R$--band
observation. A second exposure was then begun whilst the first was being
reduced. If emission lines were easily visible in the first spectrum then
the second exposure was cut short or aborted. Details of the observations
are given in Table~\ref{obstab}.

Both the imaging and spectroscopic data were reduced using standard
packages within the \iraf\ \noao\ reduction software. For the imaging
observations the bias level and dark current were first subtracted and
then the images were flat--fielded using a flat--field constructed from
all of the imaging observations. Photometry was achieved using regular
observations of photometric standards throughout the nights.

The raw spectroscopic data frames were bias subtracted, and then
flat--fielded using observations of internal calibration lamps taken with
the same instrumental set-up as the object exposures. The sky background
was removed, taking care not to include extended line emission in the sky
bands, and then the different exposures of each galaxy were combined,
where appropriate, and cosmic ray events removed. One dimensional spectra
were extracted and were wavelength calibrated using observations of CuNe
and CuAr arc lamps. Flux calibration was achieved using observations of
the spectrophotometric standard stars GD108, G60-54 and LDS749B for the
April run, and Feige-24 and L745-46A in November, and the determined
fluxes were corrected for airmass extinction.

\subsection{WHT observations}

Observations of 7 of the sources were made using LDSS2 on the William
Herschel Telescope as a backup programme during the first half of the
night of 13 August 1998. LDSS2 was used together with a SITe1 2048 by 2048
CCD, providing 0.594 arcseconds per pixel. Imaging observations were made
using the broadband Harris-R filter. Spectroscopic observations were taken
using the standard long slit with a projected size of 1.4 by 225
arcseconds, and the `medium-blue' grism, providing a scale of 5.3\AA\ per
pixel and a spectral resolution of about 13\AA.
   
Spectroscopic observations of the sources 1643$-$223 and 1920$-$077 were
made using the duel--beam ISIS spectrograph on the William Herschel
Telescope during morning twilights of 20 and 21 March 1999
respectively. In the blue arm of the spectrograph the R158B grating was
used together with a 2096 by 4200 EEV CCD, providing a spectral coverage
from the atmospheric cutoff through to longward of the dichroic at
5400\AA, and a spectral resolution of 11\AA. In the red arm the R316R
grating was used together with a 1024 by 1024 Tek CCD, providing a
spectral resolution of about 5\AA\ and a spectral range of 1525\AA. This
range was centred on 7919\AA\ for 1643$-$223 and 8245\AA\ for 1920$-$077.

Details of the observations are given in Table~\ref{obstab}. The
procedures for both the observations and the reduction of the LDSS2 data
and the ISIS data mirrored those described for the EFOSC2 observations,
except that the default spectral slit orientation for the LDSS data was
north--south. Kopff 27 was used for spectrophotometric calibration of the
LDSS data, and g193$-$74 for the ISIS data.

\begin{table*}
\caption{\label{obstab} Details of the radio, optical imaging and
spectroscopic observations.}
\begin{tabular}{cccccccccccccc}
Source   &\multicolumn{3}{c}{.................Radio.................}&\multicolumn{5}{c}{.....................Optical Imaging.....................}& 
\multicolumn{4}{c}{....................Spectroscopy....................} & Notes$^a$ \\
         & Obs. Date& Array & Freq. & Run$^b$ & Exp. Time & 
\multicolumn{2}{c}{$R$--magnitude} & ID & Slit PA & Exp. Time & $z$ & $\delta z$ & \\ 
   &   &  &[GHz]&No.&[s] & 4$''$ & 63.9kpc$^c$&    &[deg]&  [s] &       &       &  \\
0016$-$129 & 19/06/98 & BnA & 5 & 3 & 180 & 22.81 & 22.68 & RG &  0  & 1500 & 1.589 & 0.001 &   \\
0056$-$172 & Archive  & BnA &1.4& 3 &  60 & 21.11 &  ---  & RG & 315 &  900 & 1.019 & 0.002 & N \\
0101$-$128 & 19/06/98 & BnA & 5 & 3 &  60 & 20.05 &  ---  & RG &  0  &  420 & 0.387 & 0.001 & \\
0125$-$143 & 19/06/98 & BnA & 5 & 3 &  60 & 19.14 & 18.67 & RG & 270 &  800 & 0.372 & 0.001 & N \\ 
0128$-$264 & Archive  &  A  &1.4& 3 & 180 & 22.58 & 22.19 & RG & 270 & 1350 & 2.348 & 0.003 & N \\ 
0132+079   & Archive  &  A  & 5 & 3 &  60 & 20.16 & 19.79 & RG &  0  &  600 & 0.499 & 0.001 & \\  
0219+082   & 19/06/98 & BnA & 5 & 3 &  60 & 18.63 & 17.98 & RG & 270 &  300 & 0.266 & 0.001 & \\  
0310$-$150 & Archive  &  B  &1.4& 3 & 240 & 21.74 & 21.71 & RG & 270 & 1140 & 1.769 & 0.003 & \\  
0357$-$163 & 19/06/98 & BnA & 5 & 3 & 180 & 20.56 & 20.40 & RG & 270 &  900 & 0.584 & 0.002 & N \\  
0406$-$180 & Archive  & BnA &1.4& 3 & 120 & 19.21 & 19.12 & Q  & 270 &  900 & 0.722 & 0.001 & \\  
0519$-$208 & Archive  & BnA & 5 & 3 & 240 & 21.91 & 21.80 & RG & 270 &  840 & 1.086 & 0.002 &  \\  
0850$-$206 & Archive  &  B  &1.4& 1 & 300 & 22.07 & 21.93 & RG & 270 & 2400 & 1.337 & 0.002 & N \\ 
0851$-$142 & 17/09/97 &  C  & 8 & 1 & 300 & 22.63 & 22.41 & RG & 270 & 2400 & 1.665 & 0.002 & \\ 
1039+029   & Archive  &  A  & 5 & 1 & 300 & 21.27 & 20.96 & RG & 270 & 1800 & 0.535 & 0.002 & N \\
1059$-$010 & 17/09/97 &  C  & 8 &1/3& 300 & 24.20 &  ---  & RG & 270 & 1800 &  ---  &  ---  & N \\ 
1131$-$171 & 17/09/97 &  C  & 8 & 1 & 300 & 19.01 & 19.01 & Q  & 270 & 1200 & 1.618 & 0.002 & N \\ 
1138+015   & Archive  &  A  & 5 & 1 & 300 & 20.03 & 19.48 & RG & 270 &  900 & 0.443 & 0.002 & \\ 
1140$-$114 & Archive  &  A  &1.4& 1 & 300 & 23.01 & 22.95 & RG & 270 & 2400 & 1.935 & 0.001 & \\ 
1303+091   & Archive  &  A  & 5 & 1 & 300 & 22.12 & 22.16 & RG & 270 & 2400 & 1.409 & 0.002 & \\
1307+000   & Archive  &  C  & 5 & 1 & 300 & 19.56 &  ---  & RG & 270 &  900 & 0.419 & 0.001 & \\ 
1344$-$078 & Archive  &  B  &1.4& 1 & 300 & 19.73 & 19.30 & RG & 270 &  900 & 0.384 & 0.002 & N \\
1354+013   & Archive  &  A  & 5 & 1 & 300 & 21.79 & 21.44 & RG & 270 & 1200 & 0.819 & 0.001 & \\  
1411$-$057 & 17/09/97 &  C  & 8 & 1 & 300 & 22.39 & 22.23 & RG & 270 & 2400 & 1.094 & 0.002 & \\
1413$-$215 & 29/06/98 & BnA & 5 & 1 & 300 & 22.67 &  ---  & RG & --- & ---  & ---   &  ---  & N \\ 
1422$-$297 & Archive  &  B  &1.4& 1 & 300 & 22.55 & 22.50 & RG & 270 & 1650 & 1.632 & 0.001 & N \\
1434+036   & Archive  &  A  & 5 & 1 & 300 & 21.49 & 21.53 & Q  & 270 & 1650 & 1.438 & 0.001 & N \\ 
1436$-$167 & Archive  &  B  &1.4& 1 &  30 & 17.31 & 16.26 & RG & 270 &  240 & 0.146 & 0.001 & \\ 
1509+015   & Archive  &  A  & 5 & 1 & 300 & 21.61 & 21.25 & RG & 270 &  900 & 0.792 & 0.002 & \\
1602$-$093 & Archive  &  C  & 5 & 1 & 180 & 18.01 & 16.97 & RG & 270 & 1200 & 0.109 & 0.001 & \\ 
1602$-$174 & 29/06/98 & BnA & 5 &1/2& 300 & 21.69 & 21.40 & RG & 270 &  960 & 2.043 & 0.002 & N \\ 
1602$-$288 & 29/06/98 & BnA & 5 &1/2& 300 & 20.43 & 19.82 & RG & 240 & 1300 & 0.482 & 0.001 & N \\ 
1603+001   & Archive  &  A  & 5 & 1 &  30 & 16.47 & 14.98 & RG & 270 &  120 & 0.059 & 0.001 & \\ 
1621$-$115 & Archive  &  C  &1.4& 1 & 300 & 19.79 &  ---  & RG & 270 & 2100 & 0.375 & 0.002 & N \\ 
1628$-$268 & 29/06/98 & BnA & 5 & 2 & 180 & 19.10 & 17.92 & RG &  0  & 1500 & 0.166 & 0.003 & \\
1643+022   & Archive  &  A  & 5 & 1 &  60 & 17.02 &  ---  & RG & 270 &  300 & 0.095 & 0.001 & \\
1643$-$223 & 29/06/98 & BnA & 5 &1/4& 240 & 19.25 &  ---  & Q & 270 &  900 & 0.799 & 0.003 & \\
1649$-$062 & Archive  &  C  &1.4& 1 & 300 & 19.05 & 18.41 & RG & 270 &  300 & 0.236 & 0.002 & \\
1716+006   & Archive  &  A  & 5 & 1 & 300 & 19.98 & 19.89 & Q  & 270 & 1200 & 0.704 & 0.001 & N \\
1732$-$092 & 29/06/98 & BnA & 5 & 2 &  60 & 19.43 & 19.41 & RG &  0  &  900 & 0.317 & 0.001 & N \\
1810+046   & Archive  &  A  & 5 & 1 & 300 & 19.08 & 19.06 & Q  & 270 & 1200 & 1.083 & 0.002 & \\
1859$-$235 & 28/06/98 & BnA & 5 & 2 & 300 & 20.82 &  ---  & RG &  0  & 1200 & ---   &  ---  & N \\
1912$-$269 & 28/06/98 & BnA & 5 & 2 & 300 & 18.62 & 17.49 & RG &  0  &  960 & 0.226 & 0.001 & \\
1920$-$077 & 28/06/98 & BnA & 5 &2/4& 300 & 21.51 &  ---  & RG &  20 & 1200 & 0.648 & 0.001 & N \\
1953$-$077 & 28/06/98 & BnA & 5 &2/3& 420 & 22.90 &  ---  & RG & 340 & 1200 & ---   &  ---  & N \\
2025$-$155 & 28/06/98 & BnA & 5 & 3 & 120 & 19.86 &  ---  & Q  & 300 & 1700 & 1.500 & 0.003 & N \\
2120$-$166 & 28/06/98 & BnA & 5 & 3 & 120 & 20.93 & 20.94 & RG & 270 &  820 & 0.882 & 0.001 & \\
2128$-$208 & 28/06/98 & BnA & 5 & 3 &  90 & 19.49 & 19.46 & Q  & 270 &  960 & 1.615 & 0.004 & N \\
2318$-$166 & Archive  & BnA &1.4& 3 & 180 & 21.47 & 21.45 & RG & 270 &  900 & 1.414 & 0.001 & \\
2322$-$052 & 28/06/98 & BnA & 5 & 3 & 360 & 22.98 & 22.38 & RG & 270 & 3300 & 1.188 & 0.002 & N \\
2347$-$026 & Archive  &  A  &1.4& 3 & 180 & 22.95 & 22.74 & RG & 270 &  900 & 1.036 & 0.001 & \\
\end{tabular}
\raggedright $^a$ Notes can be found in Section~\ref{notes} for sources
marked with an N.\\ 
\raggedright $^b$ Where two run numbers are given the first is for
imaging, the second for spectroscopy. The run numbers correspond to:\\
\raggedright 1 --- EFOSC2 at the ESO 3.6m, April 21--22 1998.\\
\raggedright 2 --- LDSS2  at the WHT, August 13 1998.\\
\raggedright 3 --- EFOSC2 at the ESO 3.6m, November 20--21 1998. \\
\raggedright 4 --- ISIS at the WHT, March 20--21 1999.\\
\raggedright $^c$ For $\Omega = 1$ and $H_0 = 50$\kms\,Mpc$^{-1}$ \\
\end{table*}

\begin{figure*}
\centerline{
\psfig{file=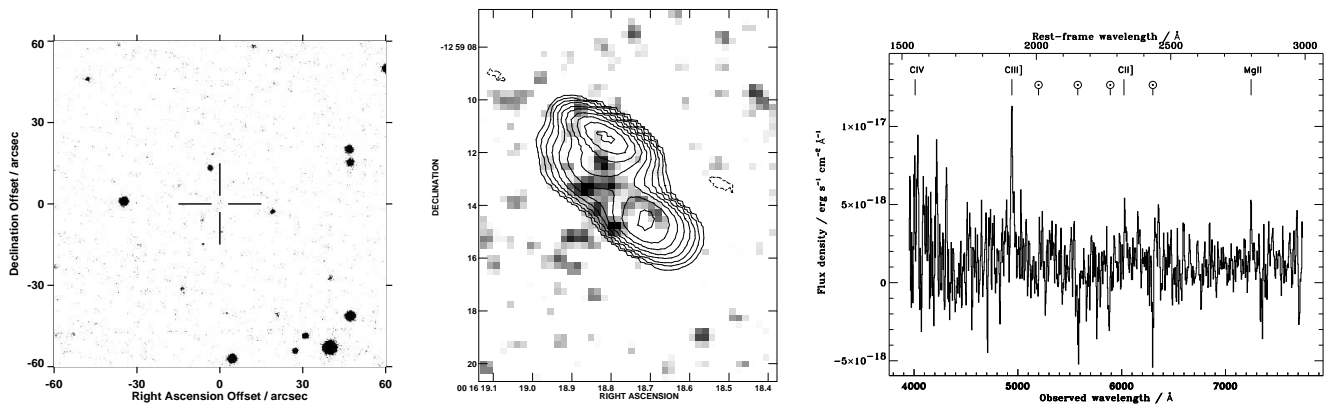,width=\textwidth,clip=}
}
\caption{(a) an $R$--band finding chart of size 120 by 120 arcseconds for
{\bf 0016--129}; (b) a zoomed view of the host galaxy, with contours of
the radio emission overlaid (the contour levels are increasing by factors
of 2 from the first contour at 0.5\,mJy/beam); (c) the spectroscopic data
with the identified emission and/or absorption lines labelled. Sky
features are indicated by an open circle with a central dot}
\end{figure*}

\begin{figure*}
\begin{center}
{\LARGE\bf VERSION INCLUDING FIGURES 2 TO 50 AVAILABLE FROM
\smallskip

http://www.strw.leidenuniv.nl/$\sim$pbest/equatpap.ps}
\end{center}
\end{figure*}

\addtocounter{figure}{49}

\section{Results of Observations}
\label{eso}

The host galaxy or quasar of the sources without pre--existing
spectroscopic redshifts has been identified from the $R$--band images in
all 50 cases. Finding charts for these identifications can be found in
Figures~1 to 50; these show a region of 2 by 2 arcminutes, centred upon
the host galaxy or quasar. The host objects are indicated by the crosses
on the finding charts. Where there may be some ambiguity as to the host
object, with more than one object lying within the radio source structure,
the justification for the labelled host is given in Section~\ref{notes} on
a source by source basis. A magnified view of the central regions can also
be found overlaid upon the radio maps in Figures~1 to 50, except for the
largest radio sources where this would simply repeat the finding chart
view.

The absolute positioning of the optical frames was determined using the
positions of between 8 and 12 unsaturated stars on the optical frames that
were also present in the APM data base \cite{mad90} or the Digitized Sky
Survey. The optical frames were registered with the sky surveys taking
account of possible rotation of the field of view, and then the precise
optical position of the host galaxy or quasar was determined. These
positions can be found in the sample collation, Table~3.  There will
remain small astrometric errors between the radio and optical images, due
to uncertainties in the absolute alignment of the radio and optical frames
of reference. The magnitude of these errors can be judged from the mean
positional difference between the optical centre of a radio galaxy or
quasar and the position of a radio core, where the latter is detected.
Unambiguous radio cores or unresolved radio sources (from a high
resolution map) are observed for 21 of the sources presented here; there
is no systematic offset between the radio core and the optical centre, and
the root--mean--squared separation is 0.55 arcseconds. This therefore is
the accuracy to which the radio and optical frames can be overlaid.  Note
that where there is an offset between the radio core position and the
position of the optical identification in our data, it is the position of
the optical source which is given in Table~3.

For 46 of the 50 galaxies a spectroscopic redshift was obtained. The
spectra are shown in Figures~1--50 with the identified lines labelled, the
resulting redshift being given in Table~\ref{obstab}. The uncertainties of
the peak positions of the measured emission lines and the variation in the
measured redshift between different lines in the spectrum were both
determined, and the larger of these values (generally the latter) was
adopted as the uncertainty on the galaxy redshift. In four cases
(1039+029, 1509+015, 1649$-$062, 2322$-$052) the redshift was based upon
only weak emission lines, or a single strong emission line together with
some weak confirming feature; notes on these sources are provided in
Section~\ref{notes}. For the four sources for which no spectroscopic
redshift has been obtained (1059$-$010, 1413$-$215, 1859$-$235,
1953$-$077), a discussion is also provided in Section~\ref{notes}. 

\subsection{Emission line properties}

Various properties of the emission lines have been determined and are
provided in Table~\ref{lineres}.  The flux of each emission line was
determined by summing the intensities of the pixels above a fitted
continuum level over a wavelength range of four times the fitted
(pre--deconvolution) full--width at half--maximum (FWHM) of the line. The
uncertainty in the measured flux of each emission line was calculated
taking account both of the measurement error from the limited
signal--to--noise of the emission line, and an uncertainty in the flux
calibration, assumed to be 10\%. Note that caution must be applied to the
use of the derived emission line fluxes, since they are measured only from
the portion of the galaxy sampled by the slit and therefore, especially at
low redshifts, are lower than the total line flux emitted by the
galaxy. These line fluxes should not be used to investigate the variation
of emission line strengths as a function of redshift. The ratios of the
line fluxes for lines not widely separated in wavelength should provide an
accurate measure, although for widely separated lines caution should again
be adopted since atmospheric diffraction means that the red and blue ends
of the spectrum might be sampling different regions of the galaxy.

Also calculated are the deconvolved FWHM of the emission lines, determined
assuming that the lines follow a Gaussian profile. The uncertainty in this
deconvolved width is a combination of the uncertainty in the measured FWHM
due to the limited signal--to--noise ratio of the line, and deconvolution
errors introduced by the uncertainty in the spectral resolution of the
observations, estimated to be 10\%. Where the fitted FWHM was found to be
less than the resolution, or the deconvolved FWHM was determined to be
less than its error, the uncertainty in the deconvolved FWHM has been
adopted as an upper limit to the deconvolved FWHM. Due to the low spectral
resolution of these observations, this was frequently the case, and little
velocity information was obtained except for the broadest lines. Finally,
the equivalent widths of the emission lines were calculated, along with
their errors. Where the equivalent width was determined to be smaller than
1.5 times its error, a value of twice the error in the equivalent width
has been adopted as an upper limit to the line equivalent width.

\subsection{Optical magnitudes and classifications}

$R$--band magnitudes were measured for the host galaxy or quasar, after
any faint objects close to the radio source host but not obviously a part
of it had been edited out of the image. The local sky level was determined
within a concentric annulus centred upon the host object with an inner
radius of between 5 and 10 arcseconds, chosen to avoid contamination from
nearby objects, and a width of 2 or 3 arcseconds. Where it was impossible
to avoid faint objects within the sky annulus, these objects were edited
out of the image before photometry was performed. In 5 cases (1307+000,
1643$-$223, 1859$-$235, 1912$-$269, 1920$-$077) it was impossible to avoid
nearby bright objects, or large numbers of faint objects, falling within
the sky annulus. In these cases the local sky level was estimated using
regions of empty sky close to the host galaxy or quasar at a range of
different position angles.

The $R$--band magnitude was determined through two different aperture
sizes. First, a 4 arcsecond diameter aperture was used, chosen to allow
comparison with the measurements of McCarthy \etal\ \shortcite{mcc96b} for
the Molonglo strip sources. Such a measurement was possible for all of the
objects, since none had significant contributions from other objects at
these radii. It should be noted, however, that this aperture measurement
seriously underestimates the luminosity of low redshift galaxies (for
quasars there is little dependence of the $R$--magnitude upon aperture
diameter). The second aperture adopted was a fixed metric aperture of
63.9\,kpc diameter; such an aperture is large enough to contain
essentially all of the light from the galaxies, the specific value of
63.9\,kpc being chosen for comparison with the measurements of Eales
\etal\ \shortcite{eal97}.  Determination of the $R$--magnitude through a
63.9\,kpc aperture was not possible in all cases: for the 4 sources
without redshifts, the aperture diameter corresponding to 63.9\,kpc is
unknown; for 8 other sources, nearby objects too bright to be accurately
edited out of the image prevented measurement of the flux out to these
radii. 

The $R$--band magnitudes of the host galaxies and quasars through these
two apertures are provided in Table~\ref{obstab}. Notice the large
differences, up to 1.5 magnitudes, between the two values for the low
redshift galaxies. The uncertainties on the measurements of these
magnitudes are about 0.1 magnitudes for $R \lta 21$, increasing to about
0.3 magnitudes by $R \approx 23$.

The nature of optical identification was classified based upon optical
appearance of the image and the presence of broad emission lines, that is,
emission lines with a deconvolved FWHM greater than 2000\kms, with an
uncertainty sufficiently low that the FWHM minus its uncertainty is above
1500\kms. If the optical image contained an unresolved component of
absolute luminosity\footnote{To calculate the K--correction needed to
derive absolute magnitudes, a power--law slope of spectral index 0.7 was
assumed for the point source continuum.} $M_{\rm R} < -24$ (roughly
equivalent to the $M_{\rm B} < -23$ limit in the quasar catalogue of
V{\'e}ron--Cetty \& V{\'e}ron 1996; cf. the discussion in Willott \etal\
1998)\nocite{ver96,wil98}, and broad emission lines, then the object was
classified as a quasar. This was true for eight of the fifty cases. Of the
remainder, 0850$-$206, 1140$-$114, 1436$-$167 are all well--resolved and
show narrow forbidden lines, but their permitted lines are broad. These
sources were therefore classified as broad--line radio
galaxies. 1732$-$092 also shows broad permitted lines, and some of its
emission comes from an unresolved component, but that component is not
sufficiently luminous to be classified as a quasar and so it also is
classified as a broad--line radio galaxy. The remainder of the
identifications are well resolved with only narrow lines and so are
classified as radio galaxies.

In Figure~\ref{rzdiag} the $R$ magnitudes are plotted against the redshift
of the source, showing the radio galaxies and quasars separately. The
$R-z$ diagram for the radio galaxies shows the well--known tight
correlation out to about a redshift of 0.8 (e.g. Eales
1985)\nocite{eal85b}, beyond which the scatter increases due to different
strengths of the alignment effect in different sources. This diagram is
powerful because the $R-z$ correlation can be used to provide supporting
evidence for redshifts determined from only weak features, and to estimate
the redshifts of the 4 remaining sources from their $R$--magnitudes. The
quasars have brighter $R$--magnitudes than would be expected from the
$R-z$ relation of the radio galaxies, due to their AGN component.

\subsection{Notes on individual sources}
\label{notes}

\noindent {\bf 0056--172:} The optical identification of this source is the
south--westerly of the two objects lying along the radio source axis; the
north--eastern object shows no strong emission lines.

\smallskip
\noindent {\bf 0125--143:} The optical identification, showing powerful
line emission, is the brighter of the two objects and is coincident with
the radio core.
 
\smallskip
\noindent {\bf 0128--264:} The identification is the faint aligned object
lying directly along the radio axis. The high background level in the
south--west is caused by a nearby bright star.

\smallskip
\noindent {\bf 0357--163:} The host galaxy appears to lie coincident with
the eastern hotspot (possibly the core if a faint radio lobe has been
missed), but this identification is secure: the galaxy shows strong
emission lines and no other object is observed within the radio source
structure.

\smallskip
\noindent {\bf 0850--206:} The host radio galaxy, showing powerful emission
lines, is the more southerly of the two objects within the radio source
structure.

\begin{figure}
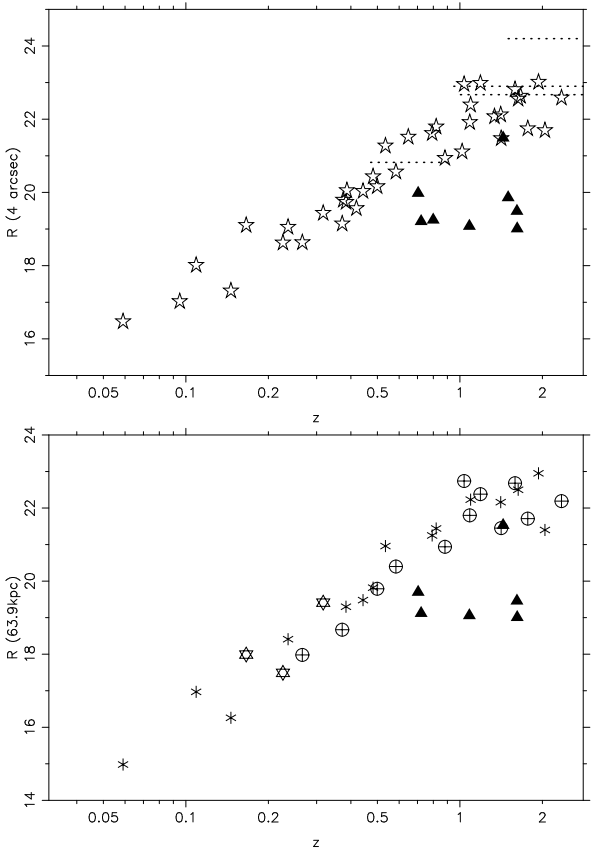

\centerline{
\psfig{file=rzdiag2.ps,angle=-90,width=7.8cm,clip=}
}
\centerline{
\psfig{file=rzdiag1.ps,angle=-90,width=7.8cm,clip=}
}
\caption{\label{rzdiag} $R$--magnitude {\it vs} redshift diagrams. The
upper plot shows the $R$--magnitude through a 4 arcsec diameter aperture,
and includes all sources: the open stars represent radio galaxies and the
filled triangles are quasars. The dotted lines depict the $R$--magnitudes
of the 4 sources without redshifts. The lower plot shows the more
physically--meaningful $R$--magnitude through a fixed 63.9\,kpc
aperture. The three types of open symbols represent the radio galaxies
observed during the three different observing runs (asterisk --- run 1;
star --- run 2; crossed circle --- run 3), showing that there were no
systematic differences between the different observing runs, and the
filled triangles are the quasars.}
\end{figure}

\smallskip
\noindent {\bf 1039+029:} The redshift of this source is based upon a
single strong emission line which, owing to its high flux, is assumed to
be [OII]~3727. This assumption is supported by the detection of a spectral
break, consistent with being at 4000\AA\ rest--frame, and by the
consistency of the $R$--magnitude of this galaxy with the $R-z$ relationship
if it is at that redshift (see Figure~\ref{rzdiag}).
 
\smallskip
\noindent {\bf 1059--010:} No spectroscopic redshift has been obtained for
this source; the only spectroscopic observations have been carried out
during twilight conditions. The proposed host galaxy is extremely faint
($R > 24$), detected at only the $\sim 4\sigma$ level, but is the only
possible identification found. The location of this galaxy at the centre
of the radio source and its elongation along the radio axis (as is common
of high redshift radio galaxies) add to its believability. Comparing the
$R$--magnitude of this galaxy with the $R-z$ relationship suggests a
minimum redshift of about 1.5.

\smallskip
\noindent {\bf 1131--171:} The northern of the two objects within the radio
source structure, a high redshift quasar, is the host identification for
this radio source.

\smallskip
\noindent {\bf 1344--078:} It is the object 2 arcseconds north of the
centre of the radio source (RA 13 44 23.60, Dec $-$07 48 25.2) which shows
strong emission lines and is therefore identified as the radio source host
galaxy.

\smallskip
\noindent {\bf 1413--215:} No spectroscopic redshift has been obtained for
this source. At the time of the imaging and spectroscopic observations,
only a low ($\sim 60$ arcsec) resolution NVSS radio map was available and
this indicated a radio position 10 arcseconds south of the true core due
to the asymmetry between the northern and southern hotspot strengths.
Given this, there was no obvious host galaxy identification, and so no
spectroscopic observations were attempted. The new high resolution VLA map
provides an unambiguous optical identification, whose $R$--band magnitude
suggests that it is above a redshift of one (cf. Figure~\ref{rzdiag}).

\smallskip
\noindent {\bf 1422--297:} The optical identification is the brighter of
the two objects towards the centre of the radio source. The fainter object
shows no strong emission lines.

\smallskip
\noindent {\bf 1434+036:} This source is classified as a quasar on account
of a sufficiently luminous unresolved optical component.  Its broad
MgII~2799 emission (FWHM $2460 \pm 408$\kms) is relatively low for
quasars, and the $R$--magnitude lies almost within the scatter of the
radio galaxy $R-z$ relation at this redshift, and so the quasar component
is likely to be not extremely powerful.

\smallskip
\noindent {\bf 1509+015:} The redshift for this galaxy is based
predominantly upon one very luminous emission line, assumed to be
[OII]~3727; only a weak emission line consistent with being MgII~2799 and
a potential 4000\AA\ break are seen to support this. Some corroborating
evidence is given by the $R$--magnitude of the galaxy, which is consistent
with the $R-z$ relationship of the sample if this redshift is correct.

\smallskip
\noindent {\bf 1602--174:} The emission line object identified as the host
galaxy of this radio source is the faint galaxy aligned along the radio
axis, coincident with the radio core.

\smallskip
\noindent {\bf 1602--288:} The two--dimensional spectrum along the radio
axis shows emission lines covering an angular extent of over 20 arcseconds
(see Figure~31), which at a redshift of $0.48$ corresponds to a spatial
extent of nearly 150\,kpc.  This emission line region extends through both
of the objects lying directly along the radio axis, centred close to the
fainter of the two objects. The continuum shape of this fainter object (RA
16 02 6.65, Dec $-$28 51 5.2) resembles an intermediate redshift radio
galaxy, whilst the brighter object (ignoring the emission lines) is
unresolved and spectroscopically a star. Therefore, although the brighter
object is coincident with what has the appearance of a radio core, it is
the fainter object which is identified as the host radio galaxy.

\smallskip
\noindent {\bf 1621--115:} The object at the centre of the radio contours
shows emission lines and so is identified as the host radio galaxy.

\smallskip
\noindent {\bf 1649--062:} The optical identification lies very close to
the centre of this large radio source. The galaxy shows only very weak
emission lines; its redshift is based upon a weak [OII]~3727 emission line
and three absorption features. The other object close to the radio source
centre is a star.

\smallskip
\noindent {\bf 1716+006:} This source is classified as a quasar since,
apart from a very close companion, its optical emission is unresolved. The
H$\gamma$ line, although weak, appears broad (FWHM of $3473 \pm
626$\kms). The $R$--band magnitude of this object is about 1.5 magnitudes
brighter than the $R-z$ relation of the radio galaxies at that redshift,
support the interpretation of a significant quasar component.

\begin{table}
\caption{\label{lineres} Emission line properties of the objects for which
spectroscopic observations were made. See text for details.}
\begin{tabular}{lrrrrrr}
Source & \multicolumn{2}{c}{Flux\,/\,10$^{-16}$} & \multicolumn{2}{c}{FWHM} &
\multicolumn{2}{c}{E.W.} \\
\& Line & \multicolumn{2}{c}{[erg/s/cm$^2$]} &
\multicolumn{2}{c}{[km\,s$^{-1}$]} & \multicolumn{2}{c}{[\AA]} \\
\\
\multicolumn{2}{l}{\bf 0016$-$129}\\
\   CIV & 1.6 &$\pm$ 0.5& $< 2985$&          & 31   &$\pm$ 14 \\
\ CIII] & 2.6 & 0.6     & $<  911$&          & 63   &      35 \\
\  CII] & 1.2 & 0.4     & $< 1657$&          &$< 64$&         \\
\  MgII & 0.6 & 0.3     & $<  525$&          &$< 32$&         \vspace*{1mm}\\
\multicolumn{2}{l}{\bf 0056$-$172}\\        
\ CIII] & 1.4 &$\pm$ 0.5& $<  559$&          & 11   &$\pm$  4 \\
\  CII] & 0.7 & 0.4     & $< 1026$&          &  5   &       2 \\
\  NeIV & 2.2 & 0.4     & $<  473$&          &    17&       3 \\
\  MgII & 4.7 & 0.6     &   1952  &$\pm$  686&    39&       5 \\
\ [NeV] &  1.2& 0.2     & $<  331$&          &    11&       2 \\
\ [NeV] &  2.5& 0.3     & $<  313$&          &    26&       4 \\
\ [OII] & 14.6& 1.5     &      803&       269&   159&      21\vspace*{1mm}\\
\multicolumn{2}{l}{\bf 0101$-$128}\\
\ [OII] & 18.2&$\pm$ 2.1&     1039&$\pm$  446&   234&$\pm$ 72 \\
\  \Hb  &  1.7& 0.4     & $<  484$&          &     8&       2 \\
\ [OIII]& 9.2& 1.0     &      702&       314&    50&       6\vspace*{1mm}\\
\multicolumn{2}{l}{\bf 0125$-$143}\\
\ [OII]& 16.2&$\pm$ 1.8&      983&$\pm$  438&    56&$\pm$  7 \\
  \Hb  &  8.6& 1.6     & $< 7770$&          &    17&       3 \\
\ [OIII]& 6.4& 0.9     & $<  458$&          &    14&       2 \\
\ [OIII]&12.4& 1.5     &      832&       393&    29&       3\vspace*{1mm}\\
\multicolumn{2}{l}{\bf 0128$-$264}\\
\  \la  &  7.9&$\pm$ 1.4&     1755&$\pm$ 1160&$<292$&         \\
\    CIV&  1.1& 0.1     &     1636&       983&   113&      23 \\
\  CIII]&  0.6& 0.6     & $< 2351$&          &$< 30$&        \vspace*{1mm}\\
\multicolumn{2}{l}{\bf 0132$+$079}\\ 	  
\ MgII  &  1.3&$\pm$ 0.9& $< 2270$&          &$< 20$&         \\
\ [NeV] &  2.4& 0.5     & $<  572$&          &    16&       3 \\
\ [OII] &  8.8& 0.9     &      592&       371&    60&       7 \\
\ [NeIII&  1.8& 0.3     & $<  409$&          &    10&       1 \\
\  \Hb  &  1.7& 0.3     & $<  352$&          &     8&       1 \\
\ [OIII]& 12.5& 1.3     & $<  278$&          &    63&       7 \\
\ [OIII]& 32.9& 3.3     & $<  264$&          &   169&      19\vspace*{1mm}\\
\multicolumn{2}{l}{\bf 0219$+$082}\\
\ [OII] & 21.8&$\pm$ 2.6& $<  465$&          &    76&$\pm$ 12 \\
\ [NeIII&  2.6& 1.1     & $<  964$&          &     7&       3 \\
\  \Hc  &  1.7& 1.0     &     1106&      1027&     2&       1 \\
\  \Hb  &  2.9& 1.2     & $<  471$&          &     3&       1 \\
\ [OIII]& 24.9& 2.6     &      829&       332&    32&       3\vspace*{1mm}\\
\multicolumn{2}{l}{\bf 0310$-$150}\\          	   	      	     	   
\   CIV & 10.2&$\pm$ 1.5&     2323&$\pm$ 1348&    84&$\pm$ 30 \\
\   HeII&  4.0& 0.8     &      810&       794&    33&      10 \\
\  CIII]&  5.0& 0.8     &     3051&      2204&    56&      13 \\
\   CII]&  2.2& 0.5     &      882&       747&    43&      16\vspace*{1mm}\\
\multicolumn{2}{l}{\bf 0357$-$163}\\  
\  MgII &  6.1&$\pm$ 1.1&     2078&$\pm$ 1493&    61&$\pm$ 15 \\
\ [NeV] &  2.2& 0.5     & $< 1511$&          &    20&       5 \\
\ [OII] &  6.7& 0.7     & $<  346$&          &    62&       8 \\
\ [NeIII&  1.9& 0.3     & $<  479$&          &    15&       3 \\
\  \Hb  &  1.4& 0.4     & $<  441$&          &     9&       2\vspace*{1mm}\\	          	   	      	     	   
\multicolumn{2}{l}{\bf 0406$-$180}\\          	   	      	     	   
\   MgII& 15.0&$\pm$ 1.8&     7126&      2775&    22&$\pm$  2 \\
\ [NeV] &  3.1& 0.4     & $<  401$&          &     4&       0 \\
\ [OII] & 14.0& 1.5     &      662&       319&    21&       2 \\
\ [NeIII&  5.4& 0.6     & $<  331$&          &     8&       1 \\
\  \Hd  &  2.0& 0.7     & $< 2329$&          &     3&       1 \\
\  \Hc  &  8.1& 1.1     &     2601&       841&    13&       1 \\
\end{tabular}
\end{table}
\addtocounter{table}{-1}

\begin{table}
\caption{{\bf cont.} Emission line properties.}
\begin{tabular}{lrrrrrr}
Source & \multicolumn{2}{c}{Flux\,/\,10$^{-16}$} & \multicolumn{2}{c}{FWHM} &
\multicolumn{2}{c}{E.W.} \\
\& Line & \multicolumn{2}{c}{[erg/s/cm$^2$]} & 
\multicolumn{2}{c}{[km\,s$^{-1}$]} & \multicolumn{2}{c}{[\AA]} \\
\\
\multicolumn{2}{l}{\bf 0519$-$208}\\          	   	      	     	   
\ CIII] &  6.0&$\pm$ 1.1&     1597&$\pm$ 1230&$<436$&         \\
\   CII]&  0.9& 0.4     & $<  787$&          &    21&      12 \\
\   MgII&  3.4& 0.6     &     2494&      1739&    84&      26 \\
\ [NeV] &  3.2& 0.5     &      707&       444&    90&      31 \\
\ [OII] & 11.7& 1.2     &      795&       275&   274&      81\vspace*{1mm}\\
\multicolumn{2}{l}{\bf 0850$-$206}\\	  	   	      	     	   
\ CIII] &  2.1&$\pm$ 0.2&     1180&$\pm$  545&    57&$\pm$  9 \\
\   CII]&  0.9& 0.1     &     2587&      1281&    22&       2 \\
\   NeIV&  1.7& 0.2     &      988&       356&    41&       4 \\
\   MgII&  1.9& 0.2     &     2384&       323&    48&       4\vspace*{1mm}\\
\multicolumn{2}{l}{\bf 0851$-$142}\\	  	   	      	     	  
\  CIV  &  3.4&$\pm$ 0.4&      961&$\pm$  554&    81&$\pm$ 20\\
\   HeII&  2.3& 0.3     & $<  536$&          &    59&      14 \\
\  CIII]&  1.6& 0.2     & $<  440$&          &    56&      13 \\
\   NeIV&  0.6& 0.1     &      907&       646&    39&      13 \\
\   MgII&  2.2& 0.4     &     6088&      5895&$<736$&        \vspace*{1mm}\\
\multicolumn{2}{l}{\bf 1039$+$029}\\	  	   	      	     	  
\ [OII]  &  1.5&$\pm$ 0.2& $<  359$&          &    30&$\pm$  3\vspace*{1mm}\\
\multicolumn{2}{l}{\bf 1131$-$171}\\	  	   	      	     	  
\  CIV  &186.4&$\pm$18.8&     7085&$\pm$  693&    49&$\pm$  5 \\
\   HeII&  6.8& 1.2     & $<  482$&          &     2&       1 \\
\  CIII]& 39.0& 4.0     &     4813&       598&    18&       1 \\
\   NeIV&  6.7& 0.8     &     8418&      3418&     3&       1\vspace*{1mm}\\
\multicolumn{2}{l}{\bf 1138$+$015}\\	  	   	      	     	  
\ [OII] & 30.5&$\pm$ 3.5&     1062&$\pm$  428&   207&$\pm$ 55 \\
\  \Hd  &  1.9& 1.6     & $< 1684$&          &$< 16$&         \\
\  \Hc  &  0.5& 2.7     & $< 7361$&          &$< 22$&         \\
\  \Hb  &  3.6& 2.6     & $< 1959$&          &$< 20$&         \\
\ [OIII]&  1.7& 0.9     & $< 9789$&          &     7&       1 \\
\ [OIII]&  3.8& 3.1     & $< 2474$&          &$< 26$&        \vspace*{1mm}\\
\multicolumn{2}{l}{\bf 1140$-$114}\\	  	   	      	     	  
\  CIV  &  1.0&$\pm$ 0.3&     2878&$\pm$ 2274&    24&$\pm$  8 \\
\   HeII&  0.5& 0.1     & $<  450$&          &    13&       3 \\
\  CIII]&  0.6& 0.1     & $<  393$&          &    15&       3 \\
\   CII]&  1.9& 0.3     &     3686&      1069&    65&      16\vspace*{1mm}\\
\multicolumn{2}{l}{\bf 1303$+$091}\\	  	   	      	     	  
\ HeII  &  2.6&$\pm$ 0.3& $<  577$&          &    21&$\pm$  3 \\
\  CIII]&  3.5& 0.4     &      981&       491&    30&       3 \\
\   NeIV&  1.0& 0.1     & $<  377$&          &    11&       1 \\
\   MgII&  2.7& 0.3     &     1232&       341&    43&       4\vspace*{1mm}\\
\multicolumn{2}{l}{\bf 1307$+$000}\\	  	   	      	     	    
\ [NeV] &  1.6&$\pm$ 0.4& $<  758$&          &     5&$\pm$  1 \\
\ [OII] & 18.9& 1.9     & $<  375$&          &    63&       6 \\
\ [NeIII&  2.8& 0.4     & $<  375$&          &     8&       1 \\
\ [OIII]&  9.0& 1.0     & $<  289$&          &    23&       2 \\
\ [OIII]& 27.9& 2.8     & $<  279$&          &    74&       7\vspace*{1mm}\\
\multicolumn{2}{l}{\bf 1344$-$078}\\	
\ [NeV] &  3.1&$\pm$ 0.6&     2873&$\pm$ 2323&    21&$\pm$  4 \\
\ [OII] & 11.4& 1.2     &      574&       393&    67&       7 \\
\  \Hc  &  0.7& 0.5     &     1688&      1455&     2&       1 \\
\  \Hb  &  2.2& 0.4     & $<  558$&          &     6&       1 \\
\ [OIII]&  4.7& 0.6     & $<  325$&          &    14&       1 \\
\ [OIII]& 15.5& 1.6     & $<  290$&          &    51&       5\vspace*{1mm}\\
\multicolumn{2}{l}{\bf 1354$+$013}\\  	   	      	     	   
\ [NeV] &  0.6&$\pm$ 0.2& $<  792$&          &    11&$\pm$  4 \\
\  [OII]&  6.4& 0.7     &      425&       307&    95&      15 \\
\ [NeIII&  0.5& 0.2     & $<  370$&          &     6&       2\vspace*{1mm}\\
\end{tabular}
\end{table}
\addtocounter{table}{-1}

\begin{table}
\caption{{\bf cont.} Emission line properties.}
\begin{tabular}{lrrrrrr}
Source & \multicolumn{2}{c}{Flux\,/\,10$^{-16}$} & \multicolumn{2}{c}{FWHM} &
\multicolumn{2}{c}{E.W.} \\
\& Line & \multicolumn{2}{c}{[erg/s/cm$^2$]} & 
\multicolumn{2}{c}{[km\,s$^{-1}$]} & \multicolumn{2}{c}{[\AA]}\\
\\
\multicolumn{2}{l}{\bf 1411$-$057}\\	  	   	      	     	   
\ CIII] &  1.1&$\pm$ 0.2& $<  781$&          &    25&$\pm$  6 \\
\   MgII&  0.5& 0.2     &     1692&$\pm$ 1481&    16&       6 \\
\  [NeV]&  0.5& 0.2     & $<  530$&          &    15&       5 \\
\  [OII]&  3.2& 0.4     & $<  266$&          &   118&      27\vspace*{1mm}\\
\multicolumn{2}{l}{\bf 1422$-$297}\\	  	   	      	     	  
\  CIV  &  4.3&$\pm$ 0.6&     1662&$\pm$  954&    61&$\pm$ 16 \\
\   HeII&  2.1& 0.4     &     1943&      1838&    30&       8 \\
\  CIII]&  1.0& 0.2     & $<  504$&          &    19&       4 \\
\   NeIV&  0.6& 0.1     &      532&       484&    13&       3\vspace*{1mm}\\
\multicolumn{2}{l}{\bf 1434$+$036}\\	  	   	      	     	  
\ HeII  &  5.5&$\pm$ 0.6&      675&$\pm$  567&    44&$\pm$  7 \\
\  CIII]&  2.7& 0.3     & $<  491$&          &    24&       3 \\
\   NeIV&  0.8& 0.1     & $<  339$&          &    10&       1 \\
\   MgII&  0.8& 0.2     &     2460&      408 &    11&       3\vspace*{1mm}\\
\multicolumn{2}{l}{\bf 1436$-$167}\\
\ [NeV] &  4.9&$\pm$ 1.0& $<  508$&          &     6&$\pm$  1 \\
\  [OII]&124.7&12.6     &      775&       468&   140&      15 \\
\ [NeIII]&15.2& 2.1     & $<  500$&          &    14&       2 \\
\  \Hc  &  2.1& 1.8     &     1467&       787&$<  6$&         \\
\  \Hb  & 22.8& 2.4     & $<  358$&          &    10&       1 \\
\ [OIII]&144.7&14.5     & $<  349$&          &    68&       6 \\
\ [OIII]&344.2&34.4     &      580&       345&   162&      16 \\
\ [OI]  & 19.5& 2.2     & $<  464$&          &     9&       1 \\
\ \Ha   &332.3&33.3     &     2180&       267&   178&      18 \\
\ [SII] &105.0&10.6     &      843&       268&    57&       5\vspace*{1mm}\\
\multicolumn{2}{l}{\bf 1509$+$015}\\	  	   	      	     	  
\ MgII  &  1.2&$\pm$ 0.3&     1426&$\pm$ 1341&    47&$\pm$ 19 \\
\ [OII] &  5.6& 0.6     &      656&       307&   108&      15\vspace*{1mm}\\
\multicolumn{2}{l}{\bf 1602$-$093}\\	  	   	      	     	   
\ [OII] &  5.8&$\pm$ 7.9& $< 3577$&          &$< 40$&        \vspace*{1mm}\\
\multicolumn{2}{l}{\bf 1602$-$174}\\	  	   	      	     	  
\  CIV  & 10.0&$\pm$ 1.1&      915&$\pm$  442&    59&$\pm$  8 \\
\   HeII&  4.8& 0.6     &      658&       456&    32&       4 \\
\  CIII]&  2.7& 0.3     & $<  367$&          &    23&       3 \\
\   NeIV&  1.0& 0.4     & $<  498$&          &    13&       6\vspace*{1mm}\\
\multicolumn{2}{l}{\bf 1602$-$288}\\	  	   	      	     	  
\ [NeV] &  2.1&$\pm$ 0.3& $<  410$&          &    19&$\pm$  2 \\
\  [OII]& 19.0& 1.9     & $<  358$&          &   189&      20 \\
\ [NeIII]& 4.3& 0.5     & $<  347$&          &    42&       4 \\
\ \Hc   &  1.6& 0.2     & $<  315$&          &    18&       2 \\
\ \Hb   &  0.8& 0.3     & $<  274$&          &    18&       2 \\
\ [OIII]& 10.8& 1.1     & $<  270$&          &   179&      25 \\
\ [OIII]& 38.4& 3.8     & $<  267$&          &   649&      93\vspace*{1mm}\\
\multicolumn{2}{l}{\bf 1603$+$001}\\
\ \Ha   & 26.0&$\pm$ 4.8&     1973&$\pm$  928&     4&$\pm$  1 \\
\ [SII] &  6.3& 2.9     & $<  853$&          &     1&       1\vspace*{1mm}\\
\multicolumn{2}{l}{\bf 1621$-$115}\\	 	   	      	     	   
\ [OII] &  3.1&$\pm$ 0.5&      658&$\pm$  539&    28&$\pm$  4\vspace*{1mm}\\
\multicolumn{2}{l}{\bf 1628$-$268}\\	 	   	      	     	   
\ [OII] &  6.2&$\pm$ 1.7& $< 3222$&          &    86&$\pm$ 32\vspace*{1mm}\\
\multicolumn{2}{l}{\bf 1643$+$022}\\	 	   	      	     	   
\ [OII] & 28.1&$\pm$ 3.8& $<  532$&          &    13&$\pm$  1 \\
\ [NeIII]&8.8 & 1.4     & $<  538$&          &     3&       1 \\
\ \Hb   &  7.5& 1.7     & $<  376$&          &     2&       1 \\
\ [OIII]&102.1& 10.3    & $<  367$&          &    27&       2 \\
\ [OIII]&242.2& 24.3    & $<  361$&          &    65&       6 \\
\ \Ha   &290.5&29.2     &     1911&       283&    86&       8 \\
\ [SII] & 53.8& 5.6     & $<  284$&          &    17&       1\vspace*{1mm}\\
\end{tabular}
\end{table}
\addtocounter{table}{-1}

\begin{table}
\caption{{\bf cont.} Emission line properties.}
\begin{tabular}{lrrrrrr}
Source & \multicolumn{2}{c}{Flux\,/\,10$^{-16}$} & \multicolumn{2}{c}{FWHM} &
\multicolumn{2}{c}{E.W.} \\
\& Line & \multicolumn{2}{c}{[erg/s/cm$^2$]} & 
\multicolumn{2}{c}{[km\,s$^{-1}$]} & \multicolumn{2}{c}{[\AA]}\\
\\
\multicolumn{2}{l}{\bf 1643$-$223}\\	 	  	       	     	  
\  CIII]& 5.9&$\pm$ 1.3&     1400&$\pm$  938&    38&$\pm$ 11 \\
\  CII] &  5.6& 1.3     &     3719&      3537&    13&       3 \\
\  MgII & 32.5& 3.9     &     6664&      2345&    76&      10\vspace*{1mm}\\
\multicolumn{2}{l}{\bf 1649$-$062}\\	 	   	      	     	   
\ [OII] &  4.9&$\pm$ 1.2& $<  853$&          &    39&$\pm$ 11\vspace*{1mm}\\
\multicolumn{2}{l}{\bf 1716$+$006}\\	 	   	      	     	   
\ [NeV] &  1.2&$\pm$ 0.3& $<  587$&          &     4&$\pm$  1 \\
\ [OII] &  4.2& 0.5     &      412&       335&    15&       1 \\
\ [NeIII]&1.9 & 0.3     & $<  536$&          &     6&       1 \\
\ \Hc   &  4.2& 0.7     &     3473&       626&    15&       2\vspace*{1mm}\\
\multicolumn{2}{l}{\bf 1732$-$092}\\	 	   	      	     	   
\ \Hb   &140.9&$\pm$14.1&     4915&$\pm$  289&    73&$\pm$  7 \\
\ [OIII]& 26.1& 2.6     &      689&        48&    14&       1 \\
\ [OIII]& 49.6& 5.0     &      642&        43&    28&       2 \\
\ \Ha   &187.0&19.0     &     5401&       553&   223&      23\vspace*{1mm}\\
\multicolumn{2}{l}{\bf 1810$+$046}\\	 	   	      	     	   
\ CIII] & 79.3&$\pm$ 8.3&     3109&$\pm$  706&    22&$\pm$  2 \\
\  NeIV &  8.7& 1.7     &     1251&       865&     1&       1 \\
\  MgII &207.0&20.9     &     6772&       661&    42&       4 \\
\ [NeV] & 11.3& 1.9     &      590&       447&     3&       1 \\
\ [OII] & 46.3& 5.1     &      749&       287&    16&       1\vspace*{1mm}\\
\multicolumn{2}{l}{\bf 1912$-$269}\\
\ [OII] &  8.8& $\pm$ 1.2&    1149&$\pm$  602&    48&$\pm$  7 \\
\ [OIII]&  3.2&  0.5     &$<  336$&          &    11&       1\vspace*{1mm}\\
\multicolumn{2}{l}{\bf 1920$-$077}\\	 	   	     	    	  
\ \Hb   &  1.4& $\pm$ 0.5& $<  314$&         &    27&$\pm$  1 \\
\ [OIII]&  5.8&  0.7     &      502&      177&   120&      67 \\
\ [OIII]& 19.2&  2.0     &      539&       72&   401&     232\vspace*{1mm}\\
\multicolumn{2}{l}{\bf 2025$-$155}\\	
\ CIII] &  3.7& $\pm$ 0.5&    2711&$\pm$ 1379&    24&$\pm$  4 \\
\  CII] &  2.6&  0.3     &    1533&       615&     8&       1 \\
\  NeIV &  2.5&  0.3     &    1624&       564&     6&       1 \\
\  MgII &  6.9&  0.9     &    6685&      4319&    16&       2\vspace*{1mm}\\
\multicolumn{2}{l}{\bf 2120$-$166}\\	 	  	       	     	  
\  CII] &  1.6& $\pm$ 0.8&$<  817$&          &    11&$\pm$  6 \\
\  MgII &  7.3&  1.9     &$< 6839$&          &    55&      16 \\
\ [OII] & 12.4&  1.3     &     728&       306&   168&      38\vspace*{1mm}\\
\multicolumn{2}{l}{\bf 2128$-$208}\\	 	  	       	     	  
\ CIV   &  4.9& $\pm$ 0.9&    2041&$\pm$ 1737&    16&$\pm$  3 \\
\  HeII &  2.6&  0.5     &$<  825$&          &     7&       1 \\
\ CIII] &  9.2&  1.1     &    4878&      1839&    21&       2\vspace*{1mm}\\
\multicolumn{2}{l}{\bf 2318$-$166}\\	 	  	       	     	  
\  HeII &  7.8&  1.0     &$<  628$&          &    24&       4 \\
\ CIII] &  6.2&  0.7     &$<  480$&          &    29&       4 \\
\  CII] &  2.0&  0.3     &    1009&       546&    12&       2 \\
\  NeIV &  1.1&  0.2     &$<  382$&          &     6&       1 \\
\  MgII &  4.4&  0.6     &    1333&       526&    38&       6\vspace*{1mm}\\
\multicolumn{2}{l}{\bf 2322$-$052}\\	 	  	       	     	  
\ CII]  &  1.8& $\pm$ 0.2&     636&$\pm$  435&   172&$\pm$ 57 \\
\  MgII &  0.8&  0.2     &    1701&      1012&$<344$&        \vspace*{1mm}\\
\multicolumn{2}{l}{\bf 2347$-$026}\\	 	  	       	    	  
\ CII]  &  0.6& $\pm$ 0.2&$<  865$&          &$<108$&         \\
\ MgII  &  1.1&  0.3     &    1327&      1221&    52&      20 \\
\ [OII] &  4.3&  0.5     &     510&       275&   161&      36 \\
\end{tabular}
\end{table}

\smallskip
\noindent {\bf 1732--092:} The optical identification is the object lying
coincident with the bright knot of radio emission. This object is mostly 
unresolved, but the unresolved component does not contain sufficient flux
for it to be classified as a quasar. Its $R$--magnitude is comparable to 
those of radio galaxies at this redshift. The galaxy shows clearly broad 
emission lines so is identified as a broad--line radio galaxy.

\smallskip
\noindent {\bf 1859--235:} The identification lies towards the centre of
the radio source and so is reasonably secure, although no spectroscopic
redshift was obtained for this source in a 20 minute spectrum (albeit
taken at a mean airmass of 1.9). The $R$--band magnitude suggest a
redshift in the range 0.3 to 0.9 (cf. Figure~\ref{rzdiag}).

\smallskip
\noindent {\bf 1920--077:} The object lying in the gap between the two
radio lobes is identified as the host galaxy on the grounds of its
location, its resolved emission (many of the objects seen in the image are
stellar), and the fact that it shows powerful line emission.

\smallskip
\noindent {\bf 1953--077:} The object close to the southern radio lobe is
a very likely candidate host galaxy, also on the basis of a detection at
the same location in a short exposure J-band image made with WHIRCAM, the
infrared imager on the WHT. A spectrum of this object taken during
twilight conditions failed to yield a redshift for the galaxy, but the
faintness of the $R$--magnitude indicates that the galaxy is above a
redshift of one.

\smallskip
\noindent {\bf 2025--155:} It is the fainter (north--eastern) of the two
objects seen close to the radio source that shows line emission and is
identified as the host object. This object is unresolved, shows a
power--law type spectrum, and has an $R$--magnitude about 3 magnitudes
brighter than the mean $R-z$ relationship and so is identified as a
quasar.

\smallskip
\noindent {\bf 2128--208:} The brighter (eastern) of the two sources is
identified as a quasar, and is the host of this radio source. The western
object shows no strong emission lines.

\smallskip
\noindent {\bf 2322--052:} The region of faint diffuse emission close to
the eastern radio lobe shows powerful line emission and is identified as
the host radio galaxy. Two emission lines are unambiguously detected with
wavelengths consistent with being CIII]~2326 and MgII~2799 at a redshift
of 1.188. Weak features consistent with [OII]~2470 and [NeV]~3426 provide
supporting evidence, although it is surprising that CII]~1909 is not
seen. The $R$--magnitude of the host galaxy is also consistent with the
proposed redshift.

\section{The complete sample}
\label{details}

\subsection{Sample Collation}

In Table~3, details are collated of all 178 radio sources in the complete
BRL sample. The right ascension and declination of the host galaxy of each
source have positional uncertainties typically well below 1 arcsecond. The
408\,MHz flux density of the radio source is taken from the MRC catalogue
value, except for radio sources of angular size larger than 100
arcseconds, in which case the Parkes Catalogue \cite{wri90} value was used
instead (see discussion in Section~\ref{giants}). The spectral index of
the source was calculated between the 408\,MHz flux density and the flux
density at 1.4\,GHz, the latter being determined from the NVSS catalogue
\cite{con98}. The radio power of each source, corrected to a rest--frame
frequency of 408\,MHz and calculated assuming $\Omega = 1$ and $H_0 =
50$\kms\,Mpc$^{-1}$, is also tabulated.

The radio galaxies and quasars were classified, where observations of
sufficient angular resolution were available for this to be done
(otherwise they are classified as `U'), into three categories: Fanaroff
and Riley (1974; hereafter FR)\nocite{fan74} classes one (`I') and two
(`II'), and sources whose emission is dominated by the radio core, either
as a core--jet or a core--halo source (`C'). For a small number of
sources, complicated radio structure prohibits an unambiguous
classification between FR\,Is and FR\,IIs; in these cases the
classification is followed by a question mark to indicate the uncertainty,
or designated as I\,/\,II. Determination of the angular size of the radio
source depended upon the source classification. In the case of the
majority FR\,II objects, the angular separation between the hotspots in
each lobe most distant from the active nucleus was measured. For FR\,I's
and core dominated sources, the determined angular size is necessarily
more arbitrary and less robust: whatever method is used the measured
angular size of an FR\,I will always be highly dependent upon the
sensitivity and frequency of the radio observations. In this paper, the
maximum angular separation within the second highest radio contour on the
quoted radio map was used.

The (heliocentric) redshifts of the host galaxies or quasars of the radio
sources are provided in the table to an accuracy of three decimal places,
unless the original measurement was not made to such precision, in which
case it is given to the accuracy quoted in the original reference;
uncertainties on the new redshifts presented in this paper can be found in
Table~\ref{obstab}. The nature of the host object is also classified in
the table. Two sources lie nearby and are a starburst galaxy (NGC253) and
a Seyfert 2 galaxy (NGC1068); the rest of the sources are classified as
either a quasar or a radio galaxy. For the sources presented in this
paper, a discussion of the classification scheme adopted has been
presented in Section~\ref{eso}. Where possible this was also applied to
the other 128 sources, but in the majority of cases it was only possible
to accept the classification determined by the authors of the original
imaging and spectroscopy papers. For some of the radio galaxies broad
lines have been detected either in our spectroscopic observations (see
Section~\ref{eso}) or in the literature, and these radio galaxies have
been sub--classified as broad--line radio galaxies (BLRG's). It should be
noted, however, that some or many of the galaxies which remain classified
under the more general `Radio Galaxy' classification may actually be
BLRG's but the spectra are of insufficient quality to determine this.

Additionally provided in Table~3 are references to the optical
identification, spectroscopic redshift, and a radio map of the galaxy. The
optical identification reference refers to the publication which first
identified the radio source host or to the first published finding chart
of the field. The spectroscopic redshift is referenced by the first
published value; for some sources later observations have confirmed this
redshift and provided a more accurate value. Where this is the case the
first publication is still given in the reference list, but the improved
value for $z$ is used. The third reference given is to a high quality
radio map or, where none exist, the best radio data available in the
literature. For sources which have been well--studied and have many radio
maps in the literature, the reference is not necessarily to the most
sensitive or the highest angular resolution data, but is simply to a map
of sufficient quality to show clearly the important features of the radio
source provided in the table.

Notes need to be added to the classifications and redshifts of three
of these sources.

\noindent {\bf 0347+052:} The NASA/IPAC Extragalactic Database (NED) gives
a redshift for this source of 0.76, but both Allington--Smith \etal\
\shortcite{all91a} and di Serego Alighieri \etal\ \shortcite{dis94c}
derive a redshift of 0.339, and so the latter value is adopted here.

\noindent {\bf 0834$-$196:} This object is classified as a galaxy in NED,
but di Serego Alighieri \etal\ \shortcite{dis94c} point out that the host
is actually an unresolved object, and they classify it as a quasar. That
classification is adopted here.

\noindent {\bf 2030$-$230:} This object is classified as a quasar by
Kapahi \etal\ \shortcite{kap98b}, but in the rest of the literature is
classified as an N--type galaxy. It is not sufficiently luminous to be
classified as a quasar here, according to the classification scheme
adopted in Section~\ref{eso}, and so falls within the radio galaxy
population. 

\subsection{Radio selection effects: missing large sources?}
\label{giants}

An issue of concern in the definition of any sample of radio sources such
as this is the possibility that giant radio sources, which may have
sufficient flux density to be above the selection limit for the complete
sample but owing to their large angular sizes may have only a low surface
brightness, are missed from the sample. A discussion of this issue for the
3CR LRL sample can be found in Bennett \shortcite{ben62} and Riley
\shortcite{ril89}.

\begin{figure}
\centerline{
\psfig{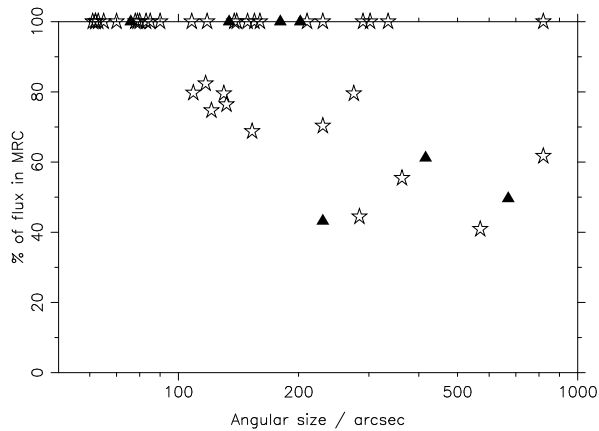} 
}
\caption{\label{fluxsize} The percentage of the flux density in the
408\,MHz Parkes Catalogue entry present in the MRC catalogue entry for all
sources of angular size larger than 60 arcsec. The sources are separated
into FR\,Is (filled triangles) and FR\,IIs (open stars).}
\end{figure}

This issue indeed turns out to be very relevant for the MRC catalogue, due
to the method of determining catalogue fluxes. As described by Large
\etal\ \shortcite{lar81}, the catalogue is based on a point-source fitting
procedure, and so the flux density of sources whose angular size is
comparable to or larger than the beam size (3 arcmins) may be
systematically underestimated by an amount which depends strongly upon the
angular structure of the source. To investigate how strong this effect is,
for all of the radio sources in the BRL sample with angular sizes in
excess of 60 arcsecs, the 408\,MHz flux density from the MRC survey has
been compared with that measured in the lower angular resolution Parkes
Catalogue \cite{wri90}; the results are shown in
Figure~\ref{fluxsize}. The flux densities determined for the MRC sources
are secure up to 100 arcsec, but above 200 arcsec the MRC flux densities
of some sources are lower by as much as a factor of two (some of this
difference may be due to the presence of other weaker sources within the
large Parkes beam, but the majority is due to an underestimate of the MRC
flux densities).  Therefore, in Table~3 the flux densities quoted for
sources larger than 100 arcsec are those taken from the Parkes Catalogue.

\begin{figure}
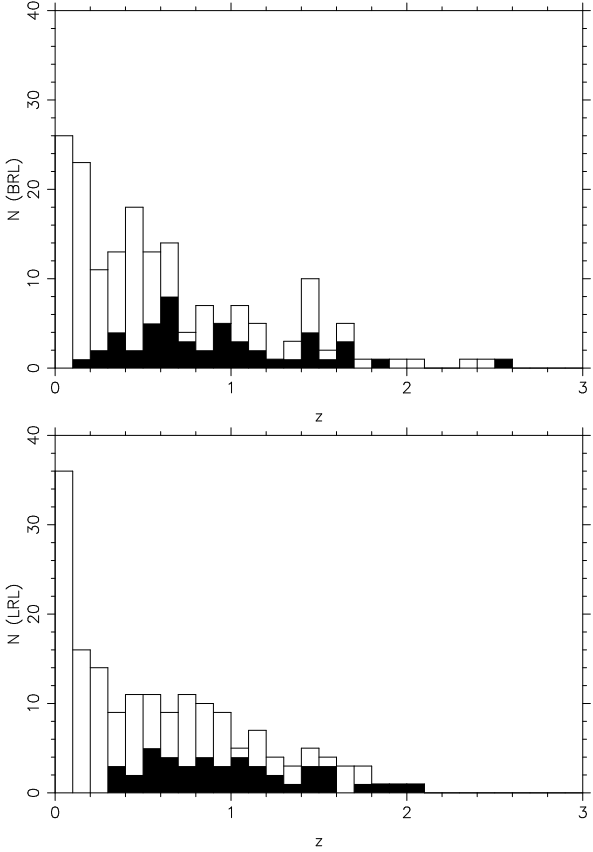

\centerline{
\psfig{file=zhistbrl.ps,angle=-90,width=7.8cm,clip=}
}
\centerline{
\psfig{file=zhistlrl.ps,angle=-90,width=7.8cm,clip=}
}
\caption{\label{zplot} Top: A histogram of the redshift distribution of
powerful radio sources in the BRL sample (shaded boxes represent
quasars). Bottom: For comparison, the equivalent histogram for the 3CR
sample of LRL.}
\end{figure}

This result raises the possibility that some sources larger than $\sim
200$ arcsec have been missed from the sample because the MRC has
underestimated their flux density, artificially placing them below the
flux density cut--off. At high redshifts this effect is likely to be of
little importance (200 arcsec corresponds to 1\,Mpc at a redshift $z
\approx 0.25$, and the 3CR sample shows that there are essentially no
higher redshift sources larger than this size of sufficient flux density),
but at $z = 0.1$ this angular size corresponds to only 500\,kpc and below
this redshift sources may be missed. Indeed, a comparison of the redshift
distributions of the BRL and LRL samples\footnote{Note: in accordance with
the new observations of Willott \etal\ (1999)\nocite{wil99}, 3C318 in the
LRL sample has been reclassified in our diagrams as a quasar of redshift
1.574 instead of a radio galaxy at $z=0.752$.}  (Figure~\ref{zplot}) shows
that the BRL sample has a slightly lower peak at $z<0.1$ which may be due
to this effect (although the combined counts at $z < 0.2$ are similar, and
a Kolmogorov--Smirnov test shows no statistically significant differences
between the two distributions). The BRL sample also contains a lower
percentage of FR\,I class sources than the LRL sample, 8\% as compared
with 16\%; FR\,Is enter the sample generally only at low redshifts and
have considerable flux which can not be well--modelled as point sources,
and so are perhaps more likely to be missed.

To summarise, whilst the sample is essentially complete at redshifts $z
\gta 0.2$, radio selection effects in the MRC catalogue may have led to a
small number of low redshift sources with angular sizes $\gta 200$\,arcsec
having been excluded from the sample.

\subsection{Sample Properties}

\begin{figure}
\centerline{
\psfig{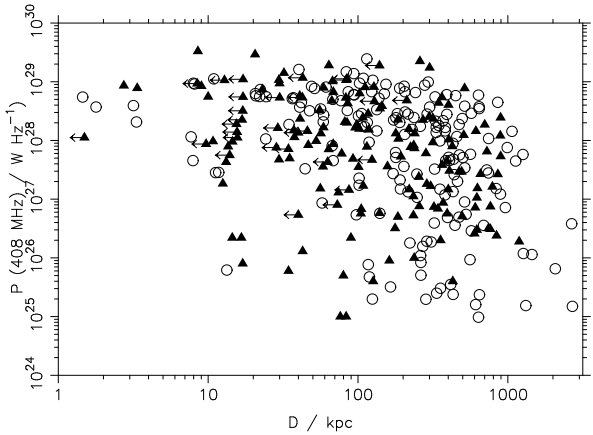} 
}
\caption{\label{pddiag} The radio power versus linear size ($P-D$)
diagram for the BRL (filled triangles) and the LRL (open circles) sources.} 
\end{figure}

\begin{figure}
\centerline{
\psfig{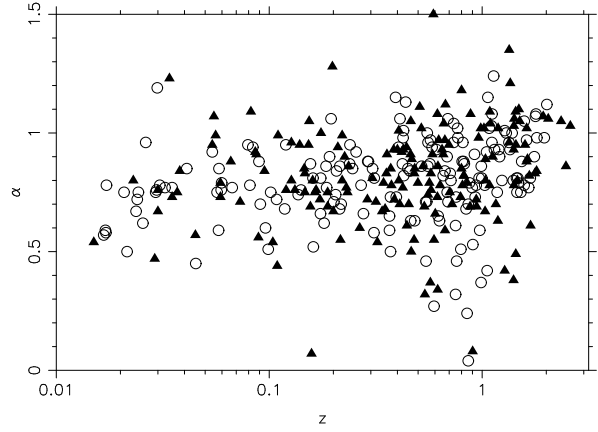} 
}
\caption{\label{zalpha} The spectral index versus redshift distribution of
the BRL (filled triangles) and LRL (open circles) sources.}
\end{figure}

\begin{figure}
\centerline{
\psfig{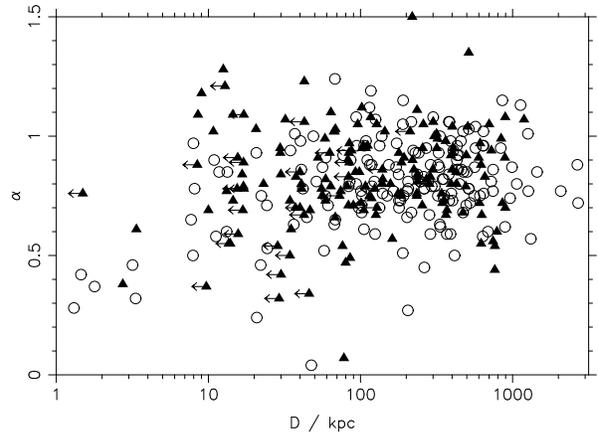} 
}
\caption{\label{dalpha} The spectral index versus linear size distribution of
the BRL (filled triangles) and LRL (open circles) sources.}
\end{figure}

Some features of the sample can be easily examined and compared to those
of the 3CR sample, to show any differences resulting from the selection at
408\,MHz instead of 178\,MHz. In Figure~\ref{pddiag} is shown the radio
power versus linear size ($P-D$) diagram for the BRL sample, and that for
the LRL sample corrected to the same rest--frame frequency. In
Figure~\ref{zalpha} the redshift versus spectral index distribution of
both samples is shown, showing the well--known increase in mean spectral
index with redshift. The two samples show very similar distributions in
both plots suggesting that there is little difference in their global
properties. The mean and standard deviation of the spectral indices of the
two samples are $\overline{\alpha_{\rm BRL}} = 0.81 \pm 0.19$ and
$\overline{\alpha_{\rm LRL}} = 0.80 \pm 0.24$, both comparable.
Figure~\ref{dalpha} shows the distribution of sources from the two samples
in the $D-\alpha$ plane; again the distributions are generally similar,
although a small excess of BRL sources is to be found with small radio
sizes ($D \lta 30$\,kpc) and steep spectra ($\alpha \gta 0.75$); it is not
entirely unexpected that the higher selection frequency will select a
larger fraction of these compact steep--spectrum sources, as synchrotron
self--absorption prohibits these from entering low frequency selected
samples. This appears to be the only important difference between the LRL
and BRL samples.

\begin{figure}
\centerline{
\psfig{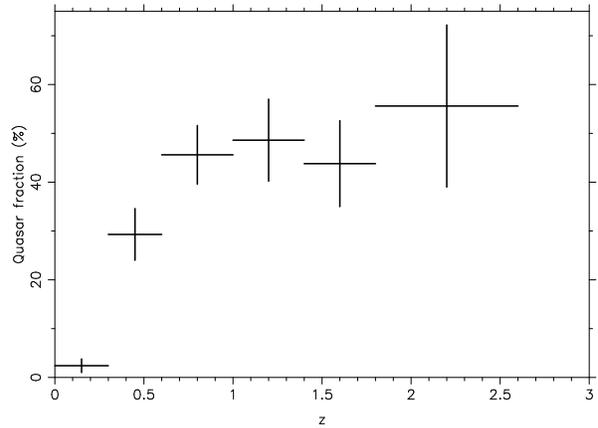}
}
\caption{\label{qfrac} The quasar fraction as a function of redshift from
the combined BRL and LRL samples. Note that inclusion of the four radio 
galaxies from the BRL sample without determined redshifts would slightly
decrease the high redshift quasar fractions, but by much less than the
size of the error bars.}
\end{figure}

\begin{figure}
\centerline{
\psfig{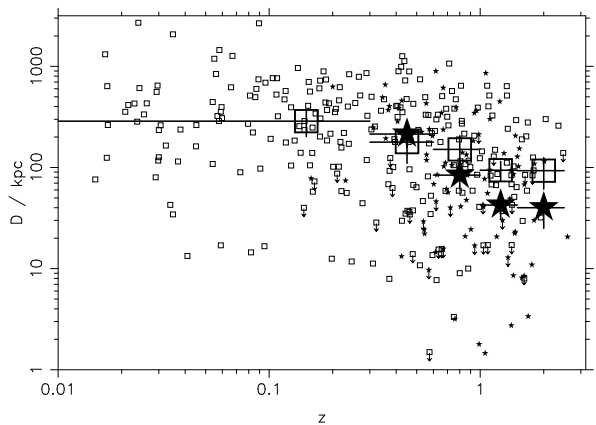} 
}
\caption{\label{dzdiag} The linear size versus redshift distribution for
the BRL and LRL samples, split into radio galaxies (open squares) and
quasars (filled stars). The larger points with error bars correspond to
the median values in five different redshift ranges. No point is plotted
for the quasars with $0 < z < 0.3$ due to the small number (3) of quasars in
this redshift range.}
\end{figure}

The redshift distribution of the two samples has already been shown to be
similar in Figure~\ref{zplot}. An interesting feature of this plot is that
the fraction of radio galaxies and quasars at redshifts $z \gta 1.5$ in
the new BRL sample remains roughly constant, indicating that the 100\%
quasar fraction beyond $z=1.8$ in the LRL sample is just due to small
number statistics. This result can be seen in Figure~\ref{qfrac} which
shows the quasar fraction as a function of redshift from the combined BRL
and LRL samples; there is no significant increase in the quasar fraction
at the highest redshifts. Both samples show a stark lack of quasars at the
lowest redshifts, which has been discussed by many authors. In
orientation--based unification schemes (e.g. Barthel 1989)\nocite{bar89}
this is partially attributed to broad--line radio galaxies being the
equivalent of the quasar population at low redshifts (e.g. see Antonucci
1993 for a review)\nocite{ant93}; more sophisticated explanations have
been proposed, including an evolution in the opening angle of the torus
with radio power (e.g. Lawrence 1991),\nocite{law91} or the presence of an
isotropic population of low excitation radio galaxies at low redshift
\cite{lai94}.

The simple unification scheme of Barthel \shortcite{bar89} makes strong
predictions for the relative linear sizes of radio galaxies and quasars;
indeed, the difference in linear sizes between the two populations in the
3CR sample was one of the factors which led him to propose the
model. Figure~\ref{dzdiag} now shows the radio size versus redshift
distribution for the BRL and LRL samples combined, plotting radio galaxies
and quasars separately. The median linear sizes of the radio galaxies and
quasars have been calculated in five separate redshift bins, the lowest
redshift bin including insufficient quasars to calculate an accurate
median. With the increased number of sources, the result of Barthel
\shortcite{bar89} still holds that radio galaxies with $0.5 < z < 1.0$
are, on average, larger than quasars in the same redshift range by about a
factor of two. This relation is also true for higher redshifts, but the
larger number statistics confirm that the result does not hold at lower
redshifts (e.g. see Singal \etal\ 1993 for the LRL sample alone; although
cf. the discussion of Section~\ref{giants} which might have a slight
effect here).  \nocite{sin93a} Again, this rules out the simplest
unification schemes, but can plausibly be explained by the modifications
discussed above (e.g. see Gopal--Krishna \etal\ 1996).\nocite{gop96}

% inclusion of large table....made elsewhere into postscript form
% owing to problems with landscape figures in mn style.

\begin{figure*}
\centerline{
\psfig{file=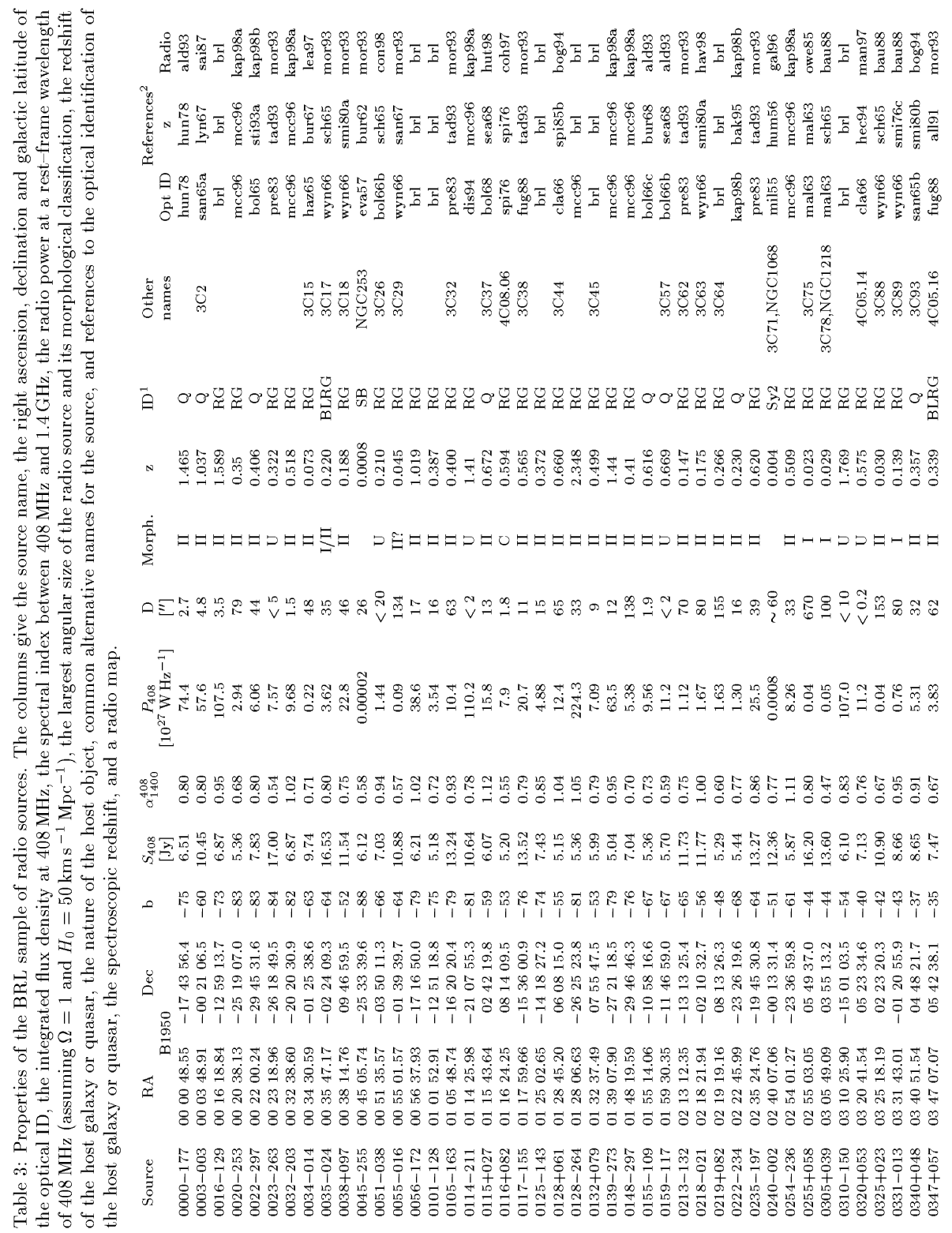,height=24cm,clip=}
}
\end{figure*}

\begin{figure*}
\centerline{
\psfig{file=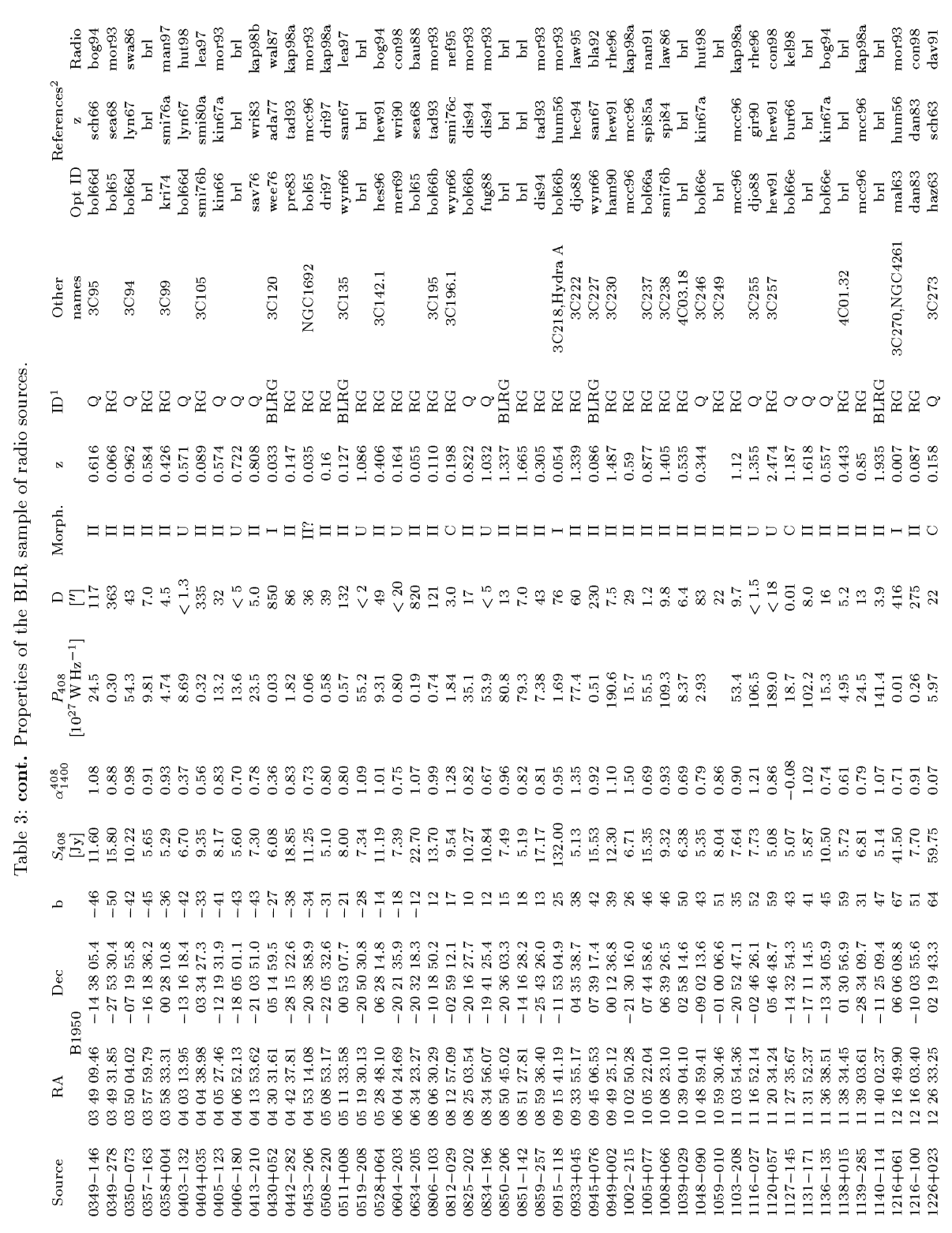,height=24cm,clip=}
}
\end{figure*}

\begin{figure*}
\centerline{
\psfig{file=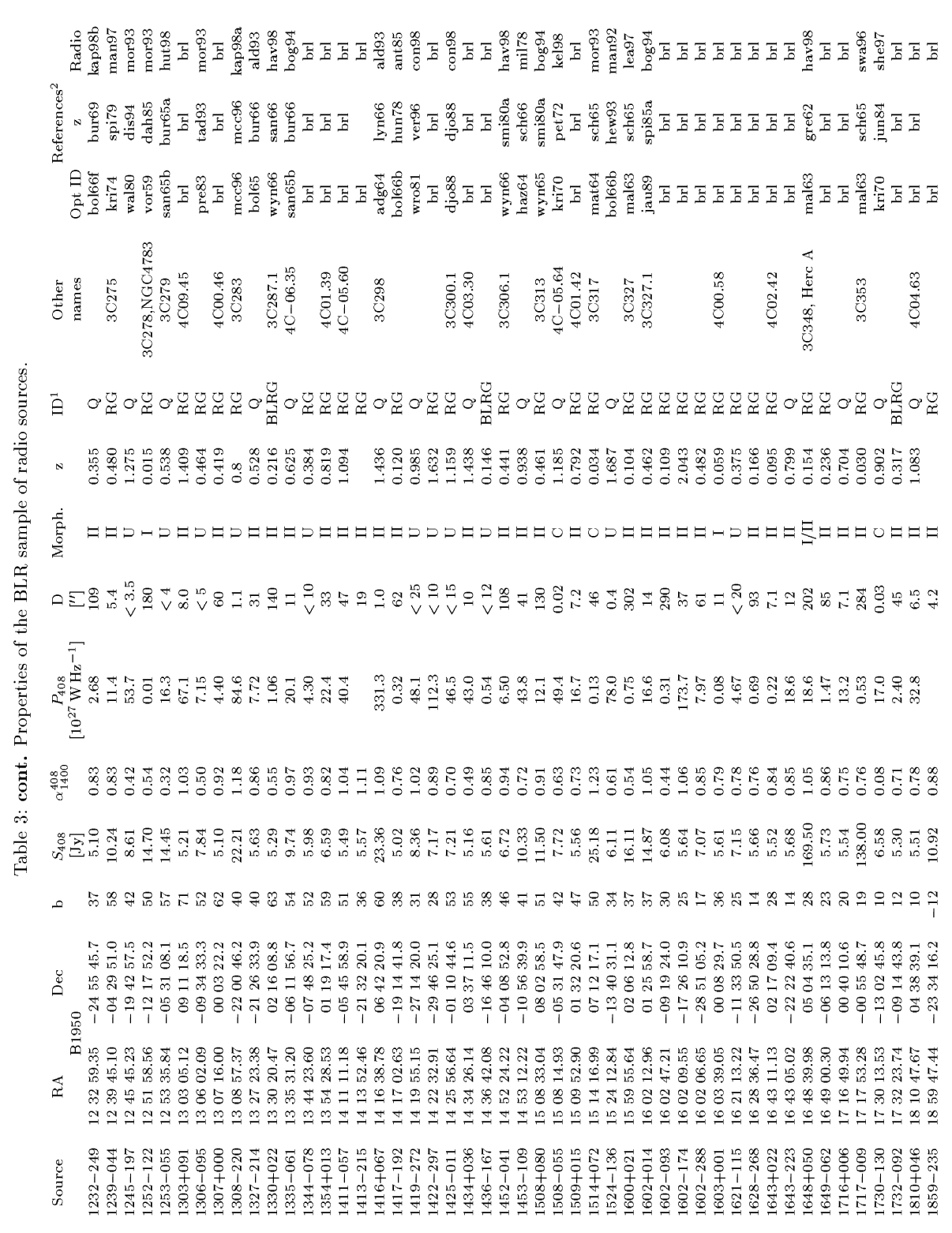,height=24cm,clip=}
}
\end{figure*}

\begin{figure*}
\centerline{
\psfig{file=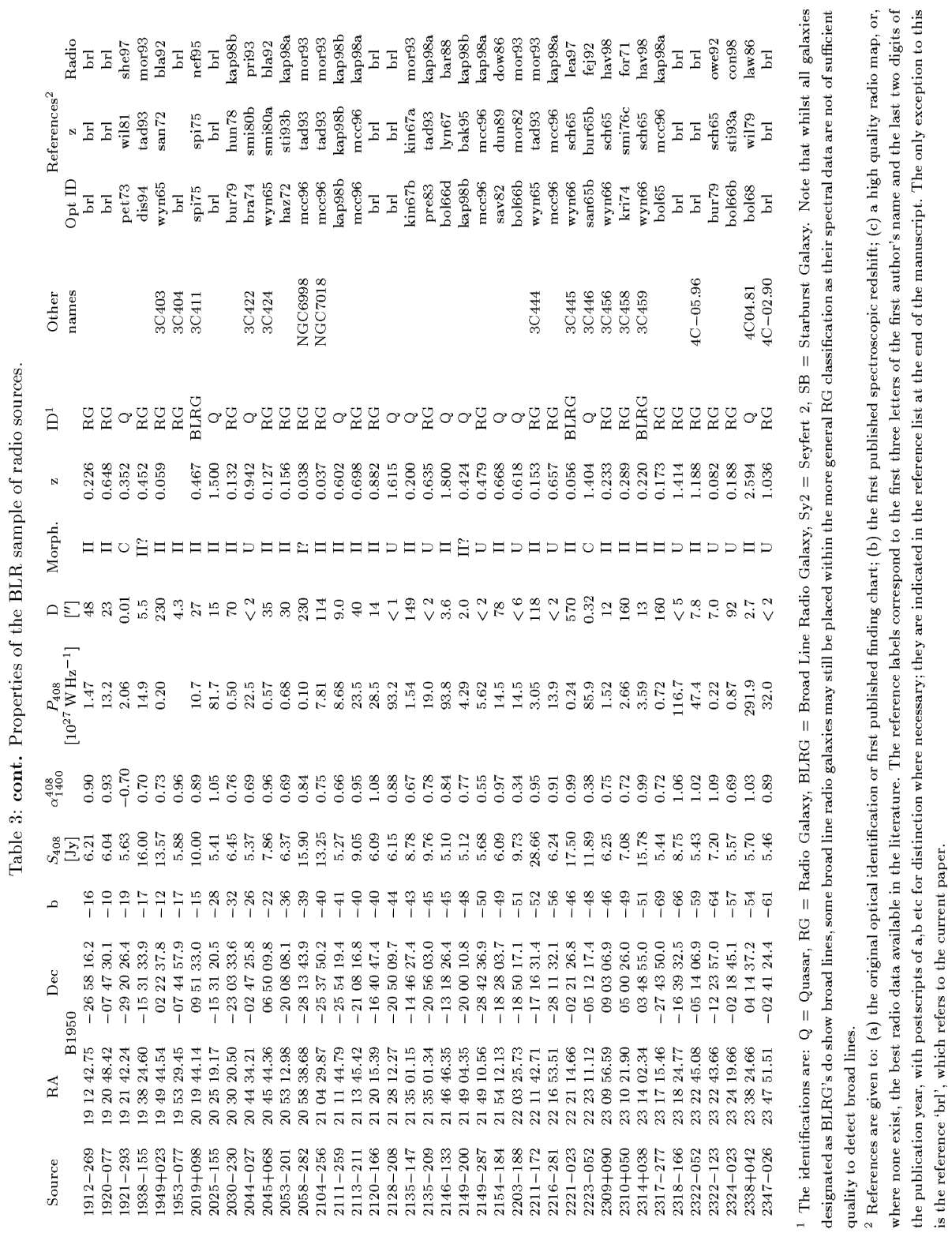,height=24cm,clip=}
}
\end{figure*}

\addtocounter{table}{1}

\nocite{ada77,adg64,ald93,all91a,ant85c,bak95,bar88,bau88,bla92,bog94,bol65}
\nocite{bol66a,bol66b,bol66c,bol66d,bol66e,bol66f,bol68,bra74,bur62,bur65a}
\nocite{bur65b,bur66b,bur67,bur68,bur69,bur79,cla66,coh97,con98,dah85,dan83}
\nocite{dav91,dis94c,djo88,dow86,dri97,dun89b,eva57,fej92,for71,fug88}
\nocite{gal96,gir90,gre62,ham90,hav98,haz63,haz64,haz65,haz72,hec94,hes96,hew91}
\nocite{hew93,hum56,hun78,hut98,jau89,jun84,kap98a,kap98b,kel98,kin66,kin67a}
\nocite{kin67b,kri70,kri74,law86,law95,lea97,lyn66,lyn67,mal63,man92,man97}
\nocite{mat64,mcc96b,mer69,mil55,mil78,mor82,mor93,nan91,nef95,owe85,owe92,pet72}
\nocite{pet73,pre83,pri93,rhe96,sai87,san65a,san65b,san66,san67,san72,sav76,sav82}
\nocite{sch63,sch65,sch66,sea68,she97b,smi76a,smi76b,smi76c,smi80a,smi80b,spi75}
\nocite{spi76,spi79,spi84b,spi85a,spi85b,sti93a,sti93b,swa86,swa96,tad93}
\nocite{ver96,vor59,wal80,wal87,wee76,wil79,wil81,wri83,wri90,wro81,wyn65,wyn66}

\section{Conclusions}
\label{conc}

Details of a new sample of the most powerful equatorial radio sources have
been collated. New radio imaging, optical imaging and spectroscopic
observations have been presented of the sources previously without
spectroscopic redshifts, leading to the complete sample being fully
optically identified and spectroscopic redshifts being available for 174 of
the 178 sources (98\%). Work to obtain the redshifts for the remaining four
sources is continuing.

Due to method of determining flux densities used for the Molonglo
Reference Catalogue, radio selection effects may have led to a small
number of radio sources subtending angular sizes larger than about 200
arcseconds being missed from the catalogue; this probably gives rise to
the slightly lower percentage of FR\,I sources in the new sample of radio
sources as compared with the revised 3CR sample. Another observed
difference is that the new sample contains a higher percentage of compact
steep spectrum sources than the 3CR sample; this was to be expected since
these sources are often missed in low frequency selected samples due to
synchrotron self--absorption. No other significant differences are found
between the properties of the new sample and those of the 3CR sample.

Due to its equatorial location and its high spectroscopic completeness,
this sample will prove very useful for studies using a combination of the
northern hemisphere instruments such as the VLA, and the new and
forthcoming southern hemisphere telescope facilities, such as the large
optical telescopes and the Atacama Large Millimetre Array.

\section*{Acknowledgements} 

This work was supported in part by the Formation and Evolution of Galaxies
network set up by the European Commission under contract ERB FMRX--
CT96--086 of its TMR programme. This work is based upon observations made
at the European Southern Observatory, La Silla, Chile, and using the
William Herschel Telescope and the Very Large Array. The William Herschel
Telescope is operated on the island of La Palma by the Isaac Newton Group
in the Spanish Observatorio del Roches de los Muchachos of the Instituto
de Astrofisica de Canarias. The National Radio Astronomy Observatory is
operated by Associated Universities Inc., under co-operative agreement
with the National Science Foundation. The Digitized Sky Surveys were
produced at the Space Telescope Science Institute under U.S. Government
grant NAG W-2166. This research has made use of the NASA/IPAC
Extragalactic Database (NED) which is operated by the Jet Propulsion
Laboratory, California Institute of Technology, under contract with the
National Aeronautics and Space Administration. The authors thank Jaron
Kurk for his work with the VLA archive data, and the referee, Steve
Rawlings, for a very careful consideration of the manuscript and a number
of helpful suggestions.

\label{lastpage}
\bibliography{pnb} 

\begin{thebibliography}{}

\bibitem[\protect\citename{Adams{\ }}{1977}]{ada77}
Adams~T.~F.,  1977, ApJ Supp., 33, 19 [ada77]

\bibitem[\protect\citename{Adgie{\ }}{1964}]{adg64}
Adgie~R.~L.,  1964, Nat, 204, 1028 [adg64]

\bibitem[\protect\citename{Aldcroft et~al.{\ }}{1993}]{ald93}
Aldcroft~T.~L.,  Elvis~M.,    Bechtold~J.,  1993, AJ, 105, 2054 [ald93]

\bibitem[\protect\citename{Allington-Smith et~al.{\ }}{1991}]{all91a}
Allington-Smith~J.~R.,  Peacock~J.~A.,    Dunlop~J.~S.,  1991, MNRAS, 253,
287 [all91]

\bibitem[\protect\citename{Antonucci{\ }}{1985}]{ant85c}
Antonucci~R.,  1985, ApJ Supp., 59, 499 [ant85]

\bibitem[\protect\citename{Antonucci{\ }}{1993}]{ant93}
Antonucci~R.,  1993, ARA\&A, 31, 473 

\bibitem[\protect\citename{Baker et~al.{\ }}{1995}]{bak95}
Baker~J.~C.,  Hunstead~R.~W.,    Brinkmann~W.,  1995, MNRAS, 277, 553 [bak95]

\bibitem[\protect\citename{Barthel{\ }}{1989}]{bar89}
Barthel~P.~D.,  1989, ApJ, 336, 606

\bibitem[\protect\citename{Barthel et~al.{\ }}{1988}]{bar88}
Barthel~P.~D.,  Miley~G.~K.,  Schilizzi~R.~T.,    Lonsdale~C.~J.,  1988, A\&A
  Supp., 73, 515 [bar88]

\bibitem[\protect\citename{Baum et~al.{\ }}{1988}]{bau88}
Baum~S.~A.,  Heckman~T.~M.,  Bridle~A.,  {van Breugel}~W. J.~M.,
  Miley~G.~K.,  1988, ApJ Supp., 68, 643 [bau88]

\bibitem[\protect\citename{Bennett{\ }}{1962}]{ben62}
Bennett~A.~S.,  1962, MemRAS, 68, 163

\bibitem[\protect\citename{Black et~al.{\ }}{1992}]{bla92}
Black~A. R.~S.,  Baum~S.~A.,  Leahy~J.~P.,  Perley~R.~A.,  Riley~J.~M.,
  Scheuer~P. A.~G.,  1992, MNRAS, 256, 186 [bla92]

\bibitem[\protect\citename{Bogers et~al.{\ }}{1994}]{bog94}
Bogers~W.~J.,  Hes~R.,  Barthel~P.~D.,    Zensus~J.~A.,  1994, A\&A Supp., 105,
  91 [bog94]

\bibitem[\protect\citename{Bolton et~al.{\ }}{1965}]{bol65}
Bolton~J.~G.,  Clarke~M.~E.,    Ekers~R.~D.,  1965, Aust. J. Phys., 18,
627 [bol65]

\bibitem[\protect\citename{Bolton \& Ekers{\ }}{1966a}]{bol66a}
Bolton~J.~G.,  Ekers~J.,  1966a, Aust. J. Phys., 19, 471 [bol66a]

\bibitem[\protect\citename{Bolton \& Ekers{\ }}{1966b}]{bol66b}
Bolton~J.~G.,  Ekers~J.,  1966b, Aust. J. Phys., 19, 559 [bol66b]

\bibitem[\protect\citename{Bolton \& Ekers{\ }}{1966c}]{bol66c}
Bolton~J.~G.,  Ekers~J.,  1966c, Aust. J. Phys., 19, 713 [bol66c]

\bibitem[\protect\citename{Bolton \& Ekers{\ }}{1966d}]{bol66f}
Bolton~J.~G.,  Ekers~J.,  1966d, Aust. J. Phys., 19, 275 [bol66f]

\bibitem[\protect\citename{Bolton \& Kinman{\ }}{1966}]{bol66e}
Bolton~J.~G.,  Kinman~T.~D.,  1966, ApJ, 145, 951 [bol66e]

\bibitem[\protect\citename{Bolton et~al.{\ }}{1966}]{bol66d}
Bolton~J.~G.,  Shimmins~A.~J.,  Ekers~J.,  Kinman~T.~D.,  Lamla~E.,
  Wirtanen~C.~A.,  1966, ApJ, 144, 1229 [bol66d]

\bibitem[\protect\citename{Bolton et~al.{\ }}{1968}]{bol68}
Bolton~J.~G.,  Shimmins~A.~J.,    Merkelijn~J.~K.,  1968, Aust. J. Phys., 21,
  81 [bol68]

\bibitem[\protect\citename{Brandie \& Bridle{\ }}{1974}]{bra74}
Brandie~G.~W.,  Bridle~A.~H.,  1974, AJ, 79, 903 [bra74]

\bibitem[\protect\citename{Burbidge{\ }}{1965}]{bur65b}
Burbidge~E.~M.,  1965, ApJ, 142, 1674 [bur65b]

\bibitem[\protect\citename{Burbidge{\ }}{1967a}]{bur67}
Burbidge~E.~M.,  1967a, ApJ, 149, L51 [bur67]

\bibitem[\protect\citename{Burbidge{\ }}{1967b}]{bur68}
Burbidge~E.~M.,  1967b, ApJ, 154, L109 [bur68]

\bibitem[\protect\citename{Burbidge et~al.{\ }}{1962}]{bur62}
Burbidge~E.~M.,  Burbidge~G.~R.,    Prendergast~K.~H.,  1962, ApJ, 136, 339
[bur62]

\bibitem[\protect\citename{Burbidge \& Kinman{\ }}{1966}]{bur66b}
Burbidge~E.~M.,  Kinman~T.~D.,  1966, ApJ, 145, 654 [bur66]

\bibitem[\protect\citename{Burbidge \& Rosenberg{\ }}{1965}]{bur65a}
Burbidge~E.~M.,  Rosenberg~F.~D.,  1965, ApJ, 142, 1673 [bur65a]

\bibitem[\protect\citename{Burbidge \& Burbidge{\ }}{1969}]{bur69}
Burbidge~G.~R.,  Burbidge~E.~M.,  1969, Nat, 222, 735 [bur69]

\bibitem[\protect\citename{Burbidge \& Crowne{\ }}{1979}]{bur79}
Burbidge~G.~R.,  Crowne~A.~H.,  1979, ApJ Supp., 40, 583 [bur79]

\bibitem[\protect\citename{Cimatti et~al.{\ }}{1996}]{cim96}
Cimatti~A.,  Dey~A.,  {van Breugel}~W.,  Antonucci~R.,    Spinrad~H.,  1996,
  ApJ, 465, 145

\bibitem[\protect\citename{Clarke et~al.{\ }}{1966}]{cla66}
Clarke~M.~E.,  Bolton~J.~G.,    Shimmins~A.~J.,  1966, Aust. J. Phys., 19,
375 [cla66]

\bibitem[\protect\citename{Cohen et~al.{\ }}{1997}]{coh97}
Cohen~M.~H.,  Vermeulen~R.~C.,  Ogle~P.~M.,  Tran~H.~D.,    Goodrich~R.~W.,
  1997, ApJ, 484, 193 [coh97]

\bibitem[\protect\citename{Condon et~al.{\ }}{1998}]{con98}
Condon~J.~J.,  Cotton~W.~D.,  Greisen~E.~W.,  Yin~Q.~F.,  Perley~R.~A.,
  Taylor~G.~B.,    Broderick~J.~J.,  1998, AJ, 115, 1693 [con98]

\bibitem[\protect\citename{Dahari{\ }}{1985}]{dah85}
Dahari~O.,  1985, ApJ Supp., 57, 643 [dah85]

\bibitem[\protect\citename{Danziger \& Goss{\ }}{1983}]{dan83}
Danziger~I.~J.,  Goss~W.~M.,  1983, MNRAS, 202, 703 [dan83]

\bibitem[\protect\citename{Davis et~al.{\ }}{1991}]{dav91}
Davis~R.~J.,  Unwin~S.~C.,    Muxlow~T. W.~B.,  1991, Nat, 354, 374 [dav91]

\bibitem[\protect\citename{Dey et~al.{\ }}{1997}]{dey97}
Dey~A.,  {van Breugel}~W. J.~M.,  Vacca~W.~D.,    Antonucci~R.,  1997, ApJ,
  490, 698

\bibitem[\protect\citename{{di Serego Alighieri} et~al.{\ }}{1994}]{dis94c}
{di Serego Alighieri}~S.,  Danziger~I.~J.,  Morganti~R.,    Tadhunter~C.~N.,
  1994, MNRAS, 269, 998 [dis94]

\bibitem[\protect\citename{Dickinson{\ }}{1997}]{dic97a}
Dickinson~M.,  1997, in Tanvir~N.~R.,  Arag{\'o}n-Salamanca~A.,   Wall~J.~V.,
  eds, HST and the high redshift Universe.
Singapore: World Scientific, p.~207

\bibitem[\protect\citename{Djorgovski et~al.{\ }}{1988}]{djo88}
Djorgovski~S.,  Spinrad~H.,  McCarthy~P.,  Dickinson~M.,  van Breugel~W. J.~M.,
     Strom~R.~G.,  1988, AJ, 96, 836 [djo88]

\bibitem[\protect\citename{Downes et~al.{\ }}{1986}]{dow86}
Downes~A. J.~B.,  Peacock~J.~A.,  Savage~A.,    Carrie~D.~R.,  1986, MNRAS,
  218, 31 [dow86]

\bibitem[\protect\citename{Drinkwater et~al.{\ }}{1997}]{dri97}
Drinkwater~M.~J.,  Webster~R.~L.,  Francis~P.~J.,  Condon~J.~J.,
  Ellison~S.~L.,  Jauncey~D.~J.,  Lovell~J.,  Peterson~B.~A.,    Savage~A.,
  1997, MNRAS, 284, 85 [dri97]

\bibitem[\protect\citename{Dunlop et~al.{\ }}{1989}]{dun89b}
Dunlop~J.~S.,  Peacock~J.~A.,  Savage~A.,  Lilly~S.~J.,  Heasley~J.~N.,
  Simon~A. J.~B.,  1989, MNRAS, 238, 1171 [dun89]

\bibitem[\protect\citename{Eales{\ }}{1985}]{eal85b}
Eales~S.~A.,  1985, MNRAS, 213, 899

\bibitem[\protect\citename{Eales et~al.{\ }}{1997}]{eal97}
Eales~S.~A.,  Rawlings~S.,  {Law--Green}~D.,  Cotter~G.,    Lacy~M.,  1997,
  MNRAS, 291, 593

\bibitem[\protect\citename{Evans{\ }}{1957}]{eva57}
Evans~D.~S.,  1957, Cape Atlas of Southern Galaxies [eva57]

\bibitem[\protect\citename{Fanaroff \& Riley{\ }}{1974}]{fan74}
Fanaroff~B.~L.,  Riley~J.~M.,  1974, MNRAS, 167, 31P

\bibitem[\protect\citename{Fejes et~al.{\ }}{1992}]{fej92}
Fejes~L.,  Porcas~R.~W.,    Akujor~C.~E.,  1992, A\&A, 257, 459 [fej92]

\bibitem[\protect\citename{Formalont{\ }}{1971}]{for71}
Formalont~E.~B.,  1971, AJ, 76, 513 [for71]

\bibitem[\protect\citename{Fugmann et~al.{\ }}{1988}]{fug88}
Fugmann~W.,  Meisenheimer~K.,    R{\"o}ser~H.-J.,  1988, A\&A Supp., 75,
173 [fug88]

\bibitem[\protect\citename{Gallimore et~al.{\ }}{1996}]{gal96}
Gallimore~J.~F.,  Baum~S.~A.,  {O'Dea}~C.~P.,    Pedlar~A.,  1996, ApJ, 458,
  136 [gal96]

\bibitem[\protect\citename{Giraud{\ }}{1990}]{gir90}
Giraud~E.,  1990, A\&A, 234, L20 [gir90]

\bibitem[\protect\citename{{Gopal--Krishna} et~al.{\ }}{1996}]{gop96}
{Gopal--Krishna} Kulkarni~V.~K.,    Wiita~P.~J.,  1996, ApJ, 463, L1

\bibitem[\protect\citename{Greenstein{\ }}{1962}]{gre62}
Greenstein~J.~L.,  1962, ApJ, 135, 679 [gre62]

\bibitem[\protect\citename{Hammer \& {Le F{\`e}vre}{\ }}{1990}]{ham90}
Hammer~F.,  {Le F{\`e}vre}~O.,  1990, ApJ, 357, 38 [ham90]

\bibitem[\protect\citename{Harvanek \& Hardcastle{\ }}{1998}]{hav98}
Harvanek~M.,  Hardcastle~M.~J.,  1998, ApJ Supp., 119, 25 [hav98]

\bibitem[\protect\citename{Hazard{\ }}{1965}]{haz65}
Hazard~C.,  1965, in Robinson~I.,  Schild~A.,   Schucking~E.~L.,  eds,
  Quasi-stellar sources and gravitational collapse.
Chicago: University of Chicago Press, p.~135 [haz65]

\bibitem[\protect\citename{Hazard{\ }}{1972}]{haz72}
Hazard~C.,  1972, ApJ, 11, L139 [haz72]

\bibitem[\protect\citename{Hazard et~al.{\ }}{1964}]{haz64}
Hazard~C.,  Mackay~M.~B.,    Nicholson~W.,  1964, Nat, 202, 227 [haz64]

\bibitem[\protect\citename{Hazard et~al.{\ }}{1963}]{haz63}
Hazard~C.,  Mackay~M.~B.,    Shimmins~A.~J.,  1963, Nat, 197, 1037 [haz63]

\bibitem[\protect\citename{Heckman et~al.{\ }}{1994}]{hec94}
Heckman~T.~M.,  {O'Dea}~C.~P.,  Baum~S.~A.,    Laurikainen~E.,  1994, ApJ, 428,
  65 [hec94]

\bibitem[\protect\citename{Hes et~al.{\ }}{1996}]{hes96}
Hes~R.,  Barthel~P.~D.,    Fosbury~R. A.~E.,  1996, A\&A, 313, 423 [hes96]

\bibitem[\protect\citename{Hewitt \& Burbidge{\ }}{1991}]{hew91}
Hewitt~A.,  Burbidge~G.,  1991, ApJ Supp., 75, 297 [hew91]

\bibitem[\protect\citename{Hewitt \& Burbidge{\ }}{1993}]{hew93}
Hewitt~A.,  Burbidge~G.,  1993, ApJ Supp., 87, 451 [hew93]

\bibitem[\protect\citename{Humason et~al.{\ }}{1956}]{hum56}
Humason~M.~L.,  Mayall~N.~U.,    Sandage~A.~R.,  1956, AJ, 61, 97 [hum56]

\bibitem[\protect\citename{Hunstead et~al.{\ }}{1978}]{hun78}
Hunstead~R.~W.,  Murdock~H.~S.,    Shobbrook~R.~R.,  1978, MNRAS, 185, 149
[hun78] 

\bibitem[\protect\citename{Hutchings et~al.{\ }}{1998}]{hut98}
Hutchings~J.~B.,  Dewey~A.,  Chaytor~D.,  Ryneveld~S.,  Gower~A.~C.,
  Ellingson~E.,  1998, PASP, 110, 111 [hut98]

\bibitem[\protect\citename{Jauncey et~al.{\ }}{1989}]{jau89}
Jauncey~D.~L.,  Savage~A.,  Morabito~D.~D.,    Preston~R.~A.,  1989, AJ,
98, 54 [jau89]

\bibitem[\protect\citename{Junkkaranen{\ }}{1984}]{jun84}
Junkkaranen~V.~T.,  1984, PASP, 96, 539 [jun84]

\bibitem[\protect\citename{Kapahi et~al.{\ }}{1998a}]{kap98b}
Kapahi~V.~K.,  Athreya~R.~M.,  Subrahmanya~C.~R.,  Baker~J.~C.,
  Hunstead~R.~W.,  McCarthy~P.~J.,    {van Breugel}~W.,  1998a, ApJ Supp., 118,
  327 [kap98b]

\bibitem[\protect\citename{Kapahi et~al.{\ }}{1998b}]{kap98a}
Kapahi~V.~K.,  Athreya~R.~M.,  {van Breugel}~W.,  McCarthy~P.~J.,
  Subrahmanya~C.~R.,  1998b, ApJ Supp., 118, 275 [kap98a]

\bibitem[\protect\citename{Kellermann et~al.{\ }}{1998}]{kel98}
Kellermann~K.~I.,  Vermeulen~R.~C.,  Zensus~J.~A.,    Cohen~M.~H.,  1998, AJ,
  115, 1295 [kel98]

\bibitem[\protect\citename{Kinman{\ }}{1966}]{kin66}
Kinman~T.~D.,  1966, ApJ, 144, 1232 [kin66]

\bibitem[\protect\citename{Kinman et~al.{\ }}{1967}]{kin67b}
Kinman~T.~D.,  Bolton~J.,  Clarke~R.,    Sandage~A.,  1967, ApJ, 147, 848 [kin67b]

\bibitem[\protect\citename{Kinman \& Burbidge{\ }}{1967}]{kin67a}
Kinman~T.~D.,  Burbidge~E.~M.,  1967, ApJ, 148, L59 [kin67a]

\bibitem[\protect\citename{Kristian \& Sandage{\ }}{1970}]{kri70}
Kristian~J.,  Sandage~A.~R.,  1970, ApJ, 162, 391 [kri70]

\bibitem[\protect\citename{Kristian et~al.{\ }}{1974}]{kri74}
Kristian~J.,  Sandage~A.~R.,    Katem~B.,  1974, ApJ, 191, 43 [kri74]

\bibitem[\protect\citename{Laing et~al.{\ }}{1994}]{lai94}
Laing~R.~A.,  Jenkins~C.~R.,  Wall~J.~V.,    Unger~S.~W.,  1994, in
  Bicknell~G.~V.,  Dopita~M.~A.,   Quinn~P.~J.,  eds, The first Stromlo
  symposium: Physics of active galaxies.
Cambridge University Press, Cambridge, p.~201

\bibitem[\protect\citename{Laing et~al.{\ }}{1983}]{lai83}
Laing~R.~A.,  Riley~J.~M.,    Longair~M.~S.,  1983, MNRAS, 204, 151 

\bibitem[\protect\citename{Large et~al.{\ }}{1981}]{lar81}
Large~M.~I.,  Mills~B.~Y.,  Little~A.~G.,  Crawford~D.~F.,    Sutton~J.~M.,
  1981, MNRAS, 194, 693

\bibitem[\protect\citename{{Law-Green} et~al.{\ }}{1995}]{law95}
{Law-Green}~J. D.~B.,  Leahy~J.~P.,  Alexander~P.,  {Allington-Smith}~J.~R.,
  {van Breugel}~W. J.~M.,  Eales~S.~A.,  Rawlings~S.~G.,    Spinrad~H.,  1995,
  MNRAS, 274, 939 [law95]

\bibitem[\protect\citename{Lawrence{\ }}{1991}]{law91}
Lawrence~A.,  1991, MNRAS, 252, 586

\bibitem[\protect\citename{Lawrence et~al.{\ }}{1986}]{law86}
Lawrence~C.~R.,  Bennett~C.~L.,  Hewitt~J.~N.,  Langston~G.~I.,  Klotz~S.~E.,
  Burke~B.~F.,    Turner~K.~C.,  1986, ApJ Supp., 61, 105 [law86]

\bibitem[\protect\citename{Leahy et~al.{\ }}{1997}]{lea97}
Leahy~J.~P.,  Black~A. R.~S.,  {Dennett-Thorpe}~J.,  Hardcastle~M.~J.,
  Komissarov~S.,  Perley~R.~A.,  Riley~J.~M.,    Scheuer~P. A.~G.,  1997,
  MNRAS, 291, 20 [lea97]

\bibitem[\protect\citename{Lilly \& Longair{\ }}{1984}]{lil84a}
Lilly~S.~J.,  Longair~M.~S.,  1984, MNRAS, 211, 833

\bibitem[\protect\citename{Lynds{\ }}{1967}]{lyn67}
Lynds~C.~R.,  1967, ApJ, 147, 837 [lyn67]

\bibitem[\protect\citename{Lynds et~al.{\ }}{1966}]{lyn66}
Lynds~C.~R.,  Hill~S.~J.,  Heere~K.,    Stockton~A.~N.,  1966, ApJ, 144,
1244 [lyn66]

\bibitem[\protect\citename{Maddox et~al.{\ }}{1990}]{mad90}
Maddox~S.~J.,  Sutherland~W.~J.,  Efstathiou~G.,    Loveday~J.,  1990, MNRAS,
  243, 692 

\bibitem[\protect\citename{Maltby et~al.{\ }}{1963}]{mal63}
Maltby~P.,  Matthews~T.~A.,    Moffet~A.~T.,  1963, ApJ, 137, 153 [mal63]

\bibitem[\protect\citename{Mantovani et~al.{\ }}{1992}]{man92}
Mantovani~F.,  Junor~W.,  Fanti~R.,  Padrielli~L.,  Browne~I. W.~A.,
  Muxlow~T. W.~B.,  1992, MNRAS, 257, 353 [man92]

\bibitem[\protect\citename{Mantovani et~al.{\ }}{1997}]{man97}
Mantovani~F.,  Junor~W.,  Fanti~R.,  Padrielli~L.,    Saikia~D.~J.,  1997, A\&A
  Supp., 125, 573 [man97]

\bibitem[\protect\citename{Matthews et~al.{\ }}{1964}]{mat64}
Matthews~T.~A.,  Morgan~W.~W.,    Schmidt~M.,  1964, ApJ, 140, 35 [mat64]

\bibitem[\protect\citename{McCarthy et~al.{\ }}{1996}]{mcc96b}
McCarthy~P.~J.,  Kapahi~V.~K.,  {van Breugel}~W.,  Persson~S.~E.,
  Athreya~R.~M.,    Subrahmanya~C.~R.,  1996, ApJ Supp., 107, 19 [mcc96]

\bibitem[\protect\citename{McCarthy et~al.{\ }}{1989}]{mcc89}
McCarthy~P.~J.,  {van Breugel}~W. J.~M.,    Spinrad~H.,  1989, AJ, 97, 36

\bibitem[\protect\citename{McCarthy et~al.{\ }}{1987}]{mcc87}
McCarthy~P.~J.,  {van Breugel}~W. J.~M.,  Spinrad~H.,    Djorgovski~S.,  1987,
  ApJ, 321, L29

\bibitem[\protect\citename{Merkelijk{\ }}{1969}]{mer69}
Merkelijk~J.~K.,  1969, Aust. J. Phys., 22, 237 [mer69]

\bibitem[\protect\citename{Miley{\ }}{1978}]{mil78}
Miley~G.~K.,  1978, A\&A Supp., 34, 129 [mil78]

\bibitem[\protect\citename{Mills{\ }}{1955}]{mil55}
Mills~B.~Y.,  1955, Aus. J. Phys., 8, 368 [mil55]

\bibitem[\protect\citename{Morganti et~al.{\ }}{1993}]{mor93}
Morganti~R.,  Killeen~N. E.~B.,    Tadhunter~C.~N.,  1993, MNRAS, 263,
1023 [mor93]

\bibitem[\protect\citename{Morton \& Tritton{\ }}{1982}]{mor82}
Morton~D.~C.,  Tritton~K.~P.,  1982, MNRAS, 198, 669 [mor82]

\bibitem[\protect\citename{{Nan Rendong} et~al.{\ }}{1991}]{nan91}
{Nan Rendong}, Schilizzi~R.~T.,  Fanti~C.,    Fanti~R.,  1991, A\&A, 252,
513 [nan91] 

\bibitem[\protect\citename{Neff et~al.{\ }}{1995}]{nef95}
Neff~S.~G.,  Roberts~L.,    Hutchings~J.~B.,  1995, ApJ Supp., 99, 349 [nef95]

\bibitem[\protect\citename{Owen et~al.{\ }}{1985}]{owe85}
Owen~F.~N.,  {O'Dea}~C.~P.,  Inoue~M.,    Eilek~J.~A.,  1985, ApJ, 294,
L85 [owe85]

\bibitem[\protect\citename{Owen et~al.{\ }}{1992}]{owe92}
Owen~F.~N.,  White~R.~A.,    Burns~J.~O.,  1992, ApJ Supp., 80, 501 [owe92]

\bibitem[\protect\citename{Peterson \& Bolton{\ }}{1972}]{pet72}
Peterson~B.~A.,  Bolton~J.~G.,  1972, ApJ, 173, L19 [pet72]

\bibitem[\protect\citename{Peterson et~al.{\ }}{1973}]{pet73}
Peterson~B.~A.,  Bolton~J.~G.,    Shimmins~A.~J.,  1973, Ap. Lett., 15,
109 [pet73]

\bibitem[\protect\citename{Prestage \& Peacock{\ }}{1983}]{pre83}
Prestage~R.~M.,  Peacock~J.~A.,  1983, MNRAS, 204, 355 [pre83]

\bibitem[\protect\citename{Price et~al.{\ }}{1993}]{pri93}
Price~R.,  Gower~A.~C.,  Hutchings~J.~B.,  Talon~S.,  Duncan~D.,    Ross~G.,
  1993, ApJ Supp., 86, 365 [pri93]

\bibitem[\protect\citename{Rhee et~al.{\ }}{1996}]{rhe96}
Rhee~G.,  Marvel~K.,  Wilson~T.,  Roland~J.,  Bremer~M.,  Jackson~N.,
  Webb~J.,  1996, ApJ Supp., 107, 175 [rhe96]

\bibitem[\protect\citename{Riley{\ }}{1989}]{ril89}
Riley~J.~M.,  1989, MNRAS, 238, 1055

\bibitem[\protect\citename{Saikia et~al.{\ }}{1987}]{sai87}
Saikia~D.~J.,  Salter~C.~J.,    Muxlow~T. W.~B.,  1987, MNRAS, 224, 911 [sai87]

\bibitem[\protect\citename{Sandage{\ }}{1966}]{san66}
Sandage~A.~R.,  1966, ApJ, 145, 1 [san66]

\bibitem[\protect\citename{Sandage{\ }}{1967}]{san67}
Sandage~A.~R.,  1967, ApJ, 150, L145 [san67]

\bibitem[\protect\citename{Sandage{\ }}{1972}]{san72}
Sandage~A.~R.,  1972, ApJ, 178, 25 [san72]

\bibitem[\protect\citename{Sandage et~al.{\ }}{1965}]{san65a}
Sandage~A.~R.,  Veron~P.,    Wyndham~J.~D.,  1965, ApJ, 142, 1307 [san65a]

\bibitem[\protect\citename{Sandage \& Wyndham{\ }}{1965}]{san65b}
Sandage~A.~R.,  Wyndham~J.~D.,  1965, ApJ, 141, 328 [san65b]

\bibitem[\protect\citename{Savage et~al.{\ }}{1982}]{sav82}
Savage~A.,  Bolton~J.~G.,    Wall~J.~V.,  1982, MNRAS, 200, 1135 [sav82]

\bibitem[\protect\citename{Savage \& Wall{\ }}{1976}]{sav76}
Savage~A.,  Wall~J.~V.,  1976, Aust. J. Phys. Ap. Supp., N.39, 39 [sav76]

\bibitem[\protect\citename{Schmidt{\ }}{1963}]{sch63} 
Schmidt~M.,  1963, Nat, 197, 1040 [sch63]

\bibitem[\protect\citename{Schmidt{\ }}{1965}]{sch65} 
Schmidt~M.,  1965, ApJ, 141, 1 [sch65]

\bibitem[\protect\citename{Schmidt{\ }}{1966}]{sch66} 
Schmidt~M.,  1966, ApJ, 144, 443 [sch66]

\bibitem[\protect\citename{Searle \& Bolton{\ }}{1968}]{sea68}
Searle~L.,  Bolton~J.~G.,  1968, ApJ, 154, L101 [sea68]

\bibitem[\protect\citename{Shen et~al.{\ }}{1997}]{she97b}
Shen~Z.-Q.,  Wan~T.-S.,  Moran~J.~M.,  Jauncey~D.~L.,  Reynolds~J.~E.,
  Tzioumis~A.~K.,  Gough~R.~G.,  Ferris~R.~H.,  Sinclair~M.~W.,  Jiang~D.-R.,
  Hong~X.-Y.,  Liang~S.-G.,  Costa~M.~E.,  Tingay~S.~J.,  McCulloch~P.~M.,
  Lovell~J. E.~J.,  King~E.~A.,  Nicolson~G.~D.,  Murphy~D.~W.,  Meier~D.~L.,
  {van Ommen}~T.~D.,  Edwards~P.~G.,    White~G.~L.,  1997, AJ, 114, 1999
[she97] 

\bibitem[\protect\citename{Singal{\ }}{1993}]{sin93a}
Singal~A.~K.,  1993, MNRAS, 262, L27

\bibitem[\protect\citename{Smith et~al.{\ }}{1976a}]{smi76b}
Smith~H.~E.,  Burbidge~E.~M.,    Spinrad~H.,  1976a, ApJ, 210, 627 [smi76b]

\bibitem[\protect\citename{Smith \& Spinrad{\ }}{1980a}]{smi80a}
Smith~H.~E.,  Spinrad~H.,  1980a, PASP, 92, 553  [smi80a]

\bibitem[\protect\citename{Smith \& Spinrad{\ }}{1980b}]{smi80b}
Smith~H.~E.,  Spinrad~H.,  1980b, ApJ, 236, 419 [smi80b]

\bibitem[\protect\citename{Smith et~al.{\ }}{1976b}]{smi76a}
Smith~H.~E.,  Spinrad~H.,    Hunstead~R.,  1976b, ApJ, 206, 345 [smi76a]

\bibitem[\protect\citename{Smith et~al.{\ }}{1976c}]{smi76c}
Smith~H.~E.,  Spinrad~H.,    Smith~E.~G.,  1976c, PASP, 88, 621 [smi76c]

\bibitem[\protect\citename{Spinrad \& Djorgovski{\ }}{1984}]{spi84b}
Spinrad~H.,  Djorgovski~S.,  1984, ApJ, 285, L49 [spi84]

\bibitem[\protect\citename{Spinrad et~al.{\ }}{1985a}]{spi85a}
Spinrad~H.,  Djorgovski~S.,  Marr~J.,    Aguilar~L.~A.,  1985a, PASP, 97,
932 [spi85a]

\bibitem[\protect\citename{Spinrad et~al.{\ }}{1985b}]{spi85b}
Spinrad~H.,  Filippenko~A.,  Wyckoff~S.,  Stocke~J.,  Wagner~M.,    Lawrie~D.,
  1985b, ApJ, 299, L7 [spi85b]

\bibitem[\protect\citename{Spinrad et~al.{\ }}{1979}]{spi79}
Spinrad~H.,  Kron~R.~G.,    Hunstead~R.,  1979, ApJ Supp., 41, 701 [spi79]

\bibitem[\protect\citename{Spinrad et~al.{\ }}{1976}]{spi76}
Spinrad~H.,  Liebert~J.,  Smith~H.~E.,    Hunstead~R.,  1976, ApJ, 206,
L79 [spi76]

\bibitem[\protect\citename{Spinrad et~al.{\ }}{1975}]{spi75}
Spinrad~H.,  Smith~H.~E.,  Hunstead~R.,    Ryle~M.,  1975, ApJ, 198, 7 [spi75]

\bibitem[\protect\citename{Stickel \& K{\"u}hr{\ }}{1993}]{sti93b}
Stickel~M.,  K{\"u}hr~H.,  1993, A\&A Supp., 100, 395 [sti93b]

\bibitem[\protect\citename{Stickel et~al.{\ }}{1993}]{sti93a}
Stickel~M.,  K{\"u}hr~H.,    Fried~J.~W.,  1993, A\&A Supp., 97, 483 [sti93a]

\bibitem[\protect\citename{Swain et~al.{\ }}{1996}]{swa96}
Swain~M.~R.,  Bridle~A.~H.,    Baum~S.~A.,  1996, in Hardee~P.~E.,
  Bridle~A.~H.,   Zensus~J.~A.,  eds, Energy transport in radio galaxies and
  quasars. [swa96]
ASP Conf. Ser. 100, San Francisco, p.~299

\bibitem[\protect\citename{Swarup et~al.{\ }}{1986}]{swa86}
Swarup~G.,  Saikia~D.~J.,  Beltrametti~M.,  Sinha~R.~P.,    Salter~C.~J.,
  1986, MNRAS, 220, 1 [swa86]

\bibitem[\protect\citename{Tadhunter et~al.{\ }}{1993}]{tad93}
Tadhunter~C.~N.,  Morganti~R.,  {di Serego Alighieri}~S.,    Fosbury~R. A.~E.,
  1993, MNRAS, 263, 999 [tad93]

\bibitem[\protect\citename{Veron-Cetty \& Veron{\ }}{1996}]{ver96}
Veron-Cetty~M.-P.,  Veron~P.,  1996, ESO Scientific Report, 17, 1 [ver96]

\bibitem[\protect\citename{{Vorontsov-Vel'yaminov}{\ }}{1959}]{vor59}
{Vorontsov-Vel'yaminov}~B.,  1959, Atlas and Catalogue of Interacting Galaxies,
  Part 1.
Moscow State Univ., Moscow [vor59]

\bibitem[\protect\citename{Walker et~al.{\ }}{1987}]{wal87}
Walker~R.~C.,  Benson~J.~M.,    Unwin~S.~C.,  1987, ApJ, 316, 546 [wal87]

\bibitem[\protect\citename{Wall \& Peacock{\ }}{1985}]{wal85}
Wall~J.~V.,  Peacock~J.~A.,  1985, MNRAS, 216, 173 

\bibitem[\protect\citename{Walter \& West{\ }}{1980}]{wal80}
Walter~H.~G.,  West~R.~M.,  1980, A\&A, 86, 1 [wal80]

\bibitem[\protect\citename{Weedman{\ }}{1976}]{wee76}
Weedman~D.~W.,  1976, ApJ, 208, 30 [wee76]

\bibitem[\protect\citename{Willott et~al.{\ }}{1998}]{wil98}
Willott~C.~J.,  Rawlings~S.,  Blundell~K.~M.,    Lacy~M.,  1998, MNRAS, 300,
  625

\bibitem[\protect\citename{Willott et~al.{\ }}{1999}]{wil99}
Willott~C.~J.,  Rawlings~S.,    Jarvis~M.,  1999, MNRAS: in press

\bibitem[\protect\citename{Wills \& Wills{\ }}{1979}]{wil79}
Wills~B.~J.,  Wills~D.,  1979, ApJ Supp., 41, 689 [wil79]

\bibitem[\protect\citename{Wills \& Wills{\ }}{1981}]{wil81}
Wills~D.,  Wills~B.~J.,  1981, Nat, 289, 384 [wil81]

\bibitem[\protect\citename{Wright \& Otrupcek{\ }}{1990}]{wri90}
Wright~A.,  Otrupcek~R.,  1990, Parkes Catalogue.
Australia Telescope National Facility [wri90]

\bibitem[\protect\citename{Wright et~al.{\ }}{1983}]{wri83}
Wright~A.~E.,  Ables~J.~G.,    Allen~D.~A.,  1983, MNRAS, 205, 793 [wri83]

\bibitem[\protect\citename{Wroblewski et~al.{\ }}{1981}]{wro81}
Wroblewski~H.,  Costa~E.,    Torres~C.,  1981, A\&A, 93, 245 [wro81]

\bibitem[\protect\citename{Wyndham{\ }}{1965}]{wyn65}
Wyndham~J.~D.,  1965, AJ, 70, 384 [wyn65]

\bibitem[\protect\citename{Wyndham{\ }}{1966}]{wyn66}
Wyndham~J.~D.,  1966, ApJ, 144, 459 [wyn66]

\end{thebibliography}
\bibliographystyle{mn} 

\end{document}